\documentstyle[prl,aps,multicol,epsf,epsfig]{revtex}
\def\mob{\mbox{\sffamily\bfseries M}}
\def\elast{\mbox{\sffamily\bfseries D}}

\def\rt{\rightarrow}

\def\brt{\longrightarrow}

\def\beq{\begin{equation}}
\def\beqar{\begin{eqnarray}}
\def\eeq{\end{equation}}
\def\eeqar{\end{eqnarray}}
\def\figone#1#2#3{\begin{figure}[h!]
\centering \leavevmode
\epsfxsize=0.75\columnwidth \epsfbox{#1}
\caption{\small #2 \label{#3}}
\end{figure}
}           
\def\figtwo#1#2#3{\begin{figure}[h!]
\centering \leavevmode
\epsfxsize=0.3 \columnwidth \epsfbox{#1}
\caption{\small #2 \label{#3}}
\end{figure}
}

\def\figfour#1#2#3#4{\begin{figure}[h!]
\centering \leavevmode
\epsfxsize=.4\columnwidth \epsfbox{#1} \hfil
\epsfxsize=.4\columnwidth \epsfbox{#2}
\caption{\small #3 \label{#4}}
\end{figure}
}          
\begin{document}
\draft
\title{Weak and strong dynamic scaling in a one-dimensional driven 
coupled-field model: Effects of kinematic waves}
\author{Dibyendu Das$^1$\cite{bydib}, Abhik Basu$^{2,3}$\cite{bypur}, 
Mustansir Barma$^1$\cite{bybarma} and Sriram Ramaswamy$^4$ \cite{byjnc}}
\address{$^1$Department of Theoretical Physics, Tata Institute of Fundamental 
Research, Homi Bhabha Road, Mumbai 400005, India,\\
$^2$Poornaprajna Institute of Scientific Research, Bangalore, India,\\
$^3$Centre for Condensed Matter Theory, Department of Physics, Indian
Institute of Science, Bangalore 560012, India}
\maketitle
\begin{abstract}
We study the coupled dynamics of the displacement fields in a one
dimensional coupled-field model for drifting crystals, first proposed
by R.Lahiri and S.Ramaswamy [{\em Phys. Rev. Lett.} {\bf 79}, 1150
(1997)].  We present some exact results for the steady state and the
current in the lattice version of the model, for a special subspace in
the parameter space, within the region where the model displays
kinematic waves.  We use these to construct the effective continuum
equations corresponding to the lattice model. These equations decouple
at the linear level in terms of the eigenmodes. We examine the
long-time, large-distance properties of the correlation functions of
the eigenmodes by using symmetry arguments, Monte Carlo simulations
and self-consistent mode coupling methods. For most parameter values,
the scaling exponents of the Kardar-Parisi-Zhang equation are
obtained. However, for certain symmetry-determined values of the
coupling constants the two eigenmodes, although nonlinearly coupled, are
characterized by two distinct dynamic exponents. We discuss the
possible application of the dynamic renormalization group in this
context.
\end{abstract}

\section{Introduction}
\label{intro}
\subsection{Background} 
\label{background}
Spatial and temporal correlations in spatially extended systems with a
conservation law or a continuous invariance are widely observed to satisfy
a scaling or homogeneity property. For instance, if the system is 
described by a single scalar field $\phi(x,t)$,
the correlation function  $C(x,t) \equiv \langle \phi(0,0)\phi(x,t)\rangle$ 
satisfies
\begin{equation}
C(x,t) \approx b^{2 \chi} C(bx, b^z t).
\end{equation}
under rescaling of space by a factor $b$.  Here $z$ is the dynamic
exponent while $\chi$ describes the spatial scaling of the field.
Equation (1) holds in the rest frame of the $\phi$-fluctuations, so that if
the system has wavelike excitations, e.g., kinematic waves
\cite{lighthill} of
moving $\phi$-fluctuations, it is necessary to perform a Galilean
shift to co-move with the wave.  The exponent $z$ then describes the dissipation
of the fluctuation wave, with a fluctuation of spatial extent $\Delta x$ having
a lifetime proportional to $(\Delta x)^z$.

Now consider the scaling properties of systems with several 
coupled fields, say $\{\phi_{\alpha}, \alpha=1 \, {\rm to}
\, N\}$, whose dynamical evolution involves inter-field couplings
both at the linear and nonlinear levels which result in propagating kinematic
waves. At the linear level, the problem requires diagonalizing the
matrix of couplings.  The eigenvectors $\psi_{\mu}$ involve linear
combinations of the $\phi_{\alpha}$ and represent modes that propagate
as independent kinematic waves. The real and imaginary parts of the
eigenvalues $c_{\mu}$ encode respectively the speeds and dampings of
the corresponding kinematic waves, and in general, differ from one
wave to another.  By performing a Galilean shift with speed $c_\mu$,
one may move to the rest frame of mode $\mu$; kinematic waves
corresponding to other modes are not stationary in this frame,
however, and these moving modes also contribute to the dissipation of
mode $\mu$ as they are coupled nonlinearly to it.  The correlation
function $C_\mu(x,t) \equiv \langle\psi_\mu(0,0)\psi_\mu(x,t)
\rangle$  is expected to satisfy
\begin{equation}
C_\mu(x,t) \approx b^{2 \chi_\mu} C(bx, b^{z_\mu} t).
\end{equation}
where $\chi_\mu$ characterizes the spatial scaling of mode $\mu$ 
and $z_\mu$ is the corresponding dynamic exponent.

The question arises: Is there a single common value $z$ which characterizes
the decay of all the modes $\mu$? When the answer is yes, as in fact it 
generally is, we say that the system obeys {\it strong dynamic scaling}.  
Considerable interest therefore attaches to exceptions to this general rule. 
Accordingly, one would like to characterize the conditions for the occurrence 
of {\it weak dynamic scaling}, when at least one $z_{\mu}$ is different 
from the rest. {\it A priori}, there are
two sets of circumstances when weak dynamic scaling may be expected:

\noindent (i) If the transformation from $\phi_{\alpha}$ to $\psi_{\mu}$,
which is designed to decouple modes at the linear level, actually succeeeds
in decoupling them for the full nonlinear problem, then evidently each
mode evolves autonomously and independent $z_{\mu}$'s may arise.
In fact, a complete decoupling at the nonlinear level does occur in the context 
of a reduced model of magnetohydrodynamic turbulence \cite{epjb}, 
and may well arise in
other problems as well. In the MHD case, however, it turns out that both
modes obey evolution equations with similar (autonomous) nonlinearites,
so that a common value of $z$ ensues. But this need not be the case for
other problems.

\noindent (ii) Consider a situation in which the evolution of 
a subset of the fields, say $\{u_{\beta}\}$,
does not involve the others $\{u_{\alpha}\}$ , while the evolution of the
set $\{u_{\alpha}\}$ does involve $\{u_{\beta}\}$. In this case, 
$z_{\beta}$ and $z_{\alpha}$ may take on  distinct values. 
Indeed, this is borne out by numerical studies of two-field
dynamics \cite{ertas1} which show that weak scaling occurs if the evolution is
autonomuous in one of the two fields, or very nearly so, in which case
strong crossover effects may be expected.

One of the principal results of this paper, which we present below, 
is a {\em third} possible scenario for weak dynamical scaling, 
where {\em neither} field is autonomous.

\subsection{Results}
\label{results}

 In this paper we examine weak and strong dynamical scaling in
a system with two coupled fields, which result in two coupled
kinematic waves characterized by mode coordinates $\psi_1$ and $\psi_2$
respectively. We work both with a lattice model involving two sets
of spins, and with the corresponding continuum equations involving
two coupled scalar fields. The analysis of the lattice model is
facilitated by showing that along certain representative loci in
parameter space, the steady state has product measure form. This allows
the current to be found, and a continuum expansion to be made, with
coefficients that explicitly involve the parameters and mean occupations
of the lattice model.  This
enables us to make direct comparisons between the results of numerical
simulations of the lattice model and analytical self-consistent calculations
for the continuum equations.

Our most interesting result is the
identification of a {\it third} set of circumstances, beyond (i) and (ii)
mentioned in Section \ref{background}, 
in which weak dynamic scaling results, despite each mode being
nonlinearly coupled to the other. This involves symmetry properties of
the kinematic waves: we find weak dynamic scaling if we
choose model parameters so that the evolution equations are
invariant under inversion of the second mode coordinate $(\psi_2 \rt
-\psi_2)$ but not under $\psi_1 \rt -\psi_1$. In that case our
numerical simulations show that $z_1=3/2$, while $z_2=2$ with multiplicative 
logarithmic
corrections. We also study the problem within a self-consistent mode coupling
calculation which shows that the different dynamic exponents arise in
a rather interesting way:
the {\em linearized} version of the problem has $z = 2$ for both fields. 
The scattering of $\psi_1$ by fluctuations in $\psi_2$ and {\it vice versa} gives 
rise to singular corrections to the diffusivity for $\psi_1$, leading to $z_1=3/2$.
The fluctuations, however, cause no singular correction to the relaxation
of $\psi_2$, leaving  $z_2 = 2$. For most other parameter values, however, 
we find the more common strong dynamic scaling with $z_1=z_2=3/2$.

The remainder of this paper is organized as follows. In Section 
\ref{continuummodel}, we 
review briefly the continuum stochastic dynamical equations of Ref.\cite{rlsr}. 
In section \ref{lattice} we  
present the lattice model, and show how the condition of pairwise balance
can be used to find the exact steady state if the transition rates satisfy
a certain relation.  We characterize changes of the symmetry of the evolution
equations as overall densities are varied, and report the
results of extensive numerical simulations which show that the
values of the dynamic exponents depend strongly on these symmetries.
In particular, we present evidence for weak dynamic 
scaling when the two kinematic waves have  different symmetries.
In section \ref{analytwds} we 
describe analytical methods, primarily a one-loop self-consistent 
treatment of the continuum stochastic PDEs, for calculating the 
exponents in the weak dynamic scaling regime. 
We also outline a dynamic renormalization group  procedure for this regime, 
discuss the difficulties that arise therein, and remind the 
reader how strong dynamic scaling is restored for generic 
values of the parameters in the model.  
We close in Section \ref{summary} with a summary. 

\section{Continuum stochastic PDEs for drifting crystals}
\label{continuummodel} 
We review very briefly here the construction of our model equations of motion;
details may be found in \cite{rlsr,lahiri2}. 
The physical system which inspired the initial work on the model 
was a lattice drifting through a dissipative medium. There are at least  
two examples of this: (i) steadily sedimenting colloidal crystals and 
(ii) a flux lattice driven, by the action of the Lorentz force of an imposed 
supercurrent, through a type II superconductor. If inertia is ignored,   
the equation of motion of the displacement field ${\bf u}({\bf r},t)$ is of the 
form velocity = mobility $\times$ force, i.e.,  
\begin{equation}
\label{eomgeneral} 
\dot{\bf u}=
\mob(\nabla {\bf u})(\elast \nabla\nabla {\bf u}+{\bf F})+{\bf \zeta},
\end{equation}
where the mobility tensor $\mob$ is allowed to depend on the lattice 
distortion $\nabla {\bf u}$, the tensor $\elast$ represents elastic forces,
$\bf F$ is the driving force, and $\bf \zeta$ a suitable noise source. 
Our results are for a highly simplified model with the same 
physics as in (\ref{eomgeneral}). This model, constructed and studied
in \cite{rlsr,lahiri2}, describes the coupled dynamics of {\em two} fields 
$u_x$ and $u_z$ (the displacements transverse to and along the drift direction 
respectively), as a function of {\em one} coordinate $x$ {\em transverse} to 
the drift direction $\hat{\bf z}$. The equations of motion are
\begin{equation}
\dot{u_x}=\lambda_{12}\partial_xu_z+\gamma_1\partial_x u_x\partial_x u_z
+D_1{\partial_x}^2u_x+f_x,
\label{lr1}
\end{equation}
\begin{equation}
\dot{u_z}=\lambda_{21}\partial_xu_x+\gamma_2(\partial_x u_x)^2+\gamma_3
(\partial_x u_z)^2+D_2{\partial_x}^2u_z+f_z,
\label{lr2}
\end{equation}
where $f_x$ and $f_z$ are zero-mean, Gaussian, spatiotemporally white 
noise sources. 
The equations are invariant under the joint operations
$x \to -x, \, u_x \to - u_x$. 
For $\lambda_{12}=\lambda_{21} = 0$, (\ref{lr1}) and (\ref{lr2}) 
reduce to the Erta\c{s}-Kardar \cite{ertas1} equations for drifting polymers 
with the larger symmetry $x \to -x$ (with or without $u_x \to -u_x$). 
The system can distinguish between up and down: there is no invariance 
under inversion of $u_z$. 
The terms in (\ref{lr1}) and (\ref{lr2}) involving first spatial derivatives  
have the following interpretation: 
a tilt ($\partial_x u_z$) produces a lateral drift (at a rate which depends 
on the density perturbations $\partial_x u_x$), while the vertical speed
depends both on compressions or dilations  
($\partial_x u_x$), as well as tilts ($\partial_x u_z$).
In this paper we shall consider only the case ${\lambda_{12}\lambda_{21}}>0$ 
in which case the dispersion relation 
\begin{equation}
\label{dispersion} 
\omega=\pm\sqrt{\lambda_{12}\lambda_{21}}q-iDq^2.
\end{equation}
holds for the linearized version of (\ref{lr1}) and (\ref{lr2}) and predicts 
travelling waves at small wavenumber $q$. 
The linearly unstable case ${\lambda_{12}\lambda_{21}}<0$ has been discussed 
extensively elsewhere \cite{lahiri2}. 
The  mode coordinates corresponding to (\ref{dispersion}) are given by
\begin{equation}
\psi_{1,2}\equiv \sqrt c u_x\pm \sqrt b u_z
\label{modecoord}
\end{equation}

As discussed in \cite{lahiri2}, the long-time, large-lengthscale 
behaviour of the PDEs (\ref{lr1}) and (\ref{lr2}) are expected to 
be the same as those of a particular two-species Ising-Kawasaki model
in which the jump rate of each species depends on the local density of 
the other.  We turn next to this discrete model and its dynamics.  

\noindent{\section {\bf The lattice model}}
\label{lattice}

The lattice model is defined in terms of two sets of variables $\{\sigma_i\}$
and $\{\tau_{i - {1 \over 2}}\}$ which reside on two interpenetrating 
sublattices with periodic boundary conditions;
the former set occupies the integer sites and the latter the half-integer 
mid-bond locations of a one-dimensional lattice with $L$ sites. 
Each $\sigma_i$ and $\tau_{i-{1 \over 2}}$ is an
Ising variable taking on values $\pm 1$.
They represent discrete versions of the density and tilt fields in
the sedimentation problem: If $\sigma_i$ is
$1$, there is a particle ($+$) at site $i$, and if $\sigma_i = -1$, there is
no particle ($-$). The variable
$\tau_{i - {1 \over 2}} = 1$ and $-1$,
implies two values $/$ and $\backslash$ of the local tilt respectively. A 
typical configuration of the full system is thus
$+\backslash -/-/+\backslash -/+/+/+\backslash -$.

Both sets of variables are conserved, i.e.   $\sum \sigma_i$ and
$\sum \tau_{i-{1 \over 2}}$  and the associated densities
$\rho_1^o=\sum (1+\sigma_i)/2L$ and $\rho_2^o=\sum (1+\tau_{i-{1 \over 2}})/2L$ 
are constant.  We consider a $\tau$-dependent local field which guides the
$\sigma$-current and {\it vice versa}. Thus, for instance,
the Kawasaki exchange dynamics of the adjacent spins $\sigma_i$ and
$\sigma_{i+1}$ occurs at a rate which depends on $\tau_{i+{1 \over 2}}$.
The moves and the corresponding rates are depicted below:

\begin{eqnarray}
\left(1\right)~~& + \backslash - ~~~\brt ~~~ - \backslash 
+&~~~~~~~~r_1
\nonumber \\
\left(2\right)~~& - \backslash +~~~ \brt~~~ + \backslash 
-&~~~~~~~~r_2
\nonumber \\
\left(3\right)~~&- / + ~~~\brt~~~ + / -&~~~~~~~~r_1 \nonumber \\
\left(4\right)~~&+ / - ~~~\brt~~~ - / +&~~~~~~~~r_2 \nonumber \\
\left(5\right)~~&/ - \backslash ~~~~\brt~~~ \backslash - 
/&~~~~~~~~p_2
\nonumber \\
\left(6\right)~~&\backslash - / ~~~\brt~~~ / - 
\backslash&~~~~~~~~p_1
\nonumber \\
\left(7\right)~~&\backslash + / ~~~\brt~~~ / + 
\backslash&~~~~~~~~p_2
\nonumber \\
\left(8\right)~~&/ + \backslash ~~~\brt~~~ \backslash + 
/&~~~~~~~~p_1
\label{moves}
\end{eqnarray}

The macroscopic behaviour of the model is determined by the relative
values of the rates; a brief review of the phases and their
characteristics appears in \cite{jnurev}.  There are two distinct
regimes separated by a nonequilibrium phase boundary as depicted in
Fig. 1. The regime $p_1 > p_2$, marked SPS in Fig. 1 was explored in detail in
\cite{lahiri2}. In this phase, the system undergoes spontaneous phase 
separation of a
particularly strong sort. Along the boundary $p_1 = p_2$, marked FDPO
in Fig. 1, the system undergoes fluctuation-dominated phase ordering
of a delicate sort, as discussed in \cite{ddmb}.  Finally, in the 
phase of the model with $p_1 < p_2$, marked KW in Fig. 1, 
there is no phase separation, and fluctuations
are transported by kinematic waves. This is the regime of interest in
this paper.

\begin{figure}[htb]
\epsfxsize=10cm
\centerline{\epsffile{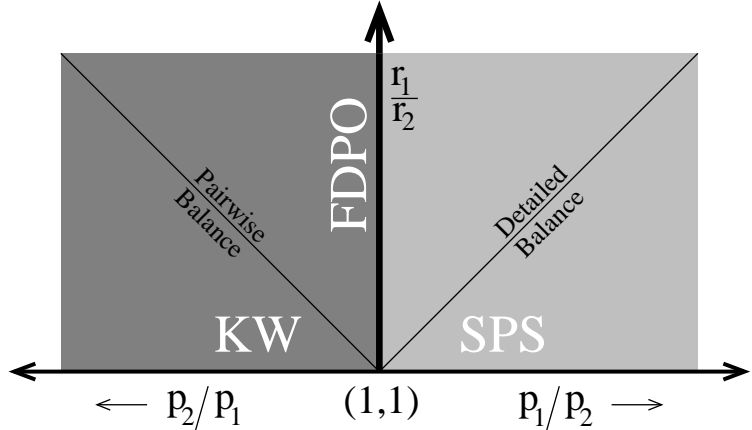}}
\caption{The phase diagram of the Lahiri-Ramaswamy model. The strongly
phase separated phase (SPS) is separated from the kinematic wave phase (KW)
of interest in this paper by the threshold line $p_1=p_2$ along which there
is fluctuation-dominated phase ordering. The steady state can be found 
exactly along the line $p_2/p_1 = r_1/r_2$,  by using the condition of 
pairwise balance.} 
\end{figure}

{\subsection {\bf Exact results for the steady state}}

The steady state can be found exactly provided that $r_1=p_2$ and $r_2=p_1$
(along the line marked Pairwise Balance in Fig. 1). To this end, let us 
choose new symbols to denote the values of the variables $\sigma$
and $\tau$: use $1$ if a site or bond is occupied by a $+$ or $/$,
and use $0$ for $-$ or $\backslash$. Then the moves (1)-(8) listed above 
reduce to moves $(a), (b), ({\bar a}), ({\bar b})$ as follows:
\beq
(1),(5)~ \Rightarrow ~(a):~1 0 0 \brt 0 0 1~~~~~~~~~ p_2
\label{newa}
\eeq
\beq
(3), (7)~ \Rightarrow~ (b):~0 1 1 \brt 1 1 0~~~~~~~~~ p_2
\label{newb}
\eeq
\beq
(2), (6)~ \Rightarrow~ ({\bar a}): ~001 \brt 100~~~~~~~~~ p_1
\label{newa-}
\eeq
\beq
(4), (8)~ \Rightarrow~ ({\bar b}):~110 \brt 011~~~~~~~~~ p_1
\label{newb-}
\eeq                                                        
The use of the new symbols $1$ and $0$ explicitly brings
out the fact that the dynamical moves on the two sublattices  are
alike for the choice of these special relations between the rates.
This is crucial for further analysis. 

In this new representation,
a configuration $C$ is specified by the occupations of all sites. The time
evolution of the probability $P(C)$ of the occurrence of $C$
is given by the master equation
\beq {dP(C) \over dt} = \sum_{C'} W({C'}
\brt C)P({C'}) - W(C \brt {C'})P(C)
\label{master}
\eeq
where the transition rates $W(C \brt C')$ are given by $p_2$ and $p_1$ for changes of 
configurations involving moves $(a)$ or $(b)$, and $ ({\bar a})$ or $({\bar b})$
respectively.

The dynamics preserves the sublattice densities  $\rho_1^o$ and $\rho_2^o$.
Within the subspace labelled by $(\rho_1^o,\rho_2^o)$,
one can see that the system is ergodic by noting that the dynamics induces
a leftward drift of a pairs $00$ and $11$ with rate
$p_2$, and a rightward drift of such pairs with rate $p_1$. By successive
application of  moves (a), (b) in Eqs. (\ref{newa}), (\ref{newb}) and their 
reverses,
any configuration $C$ in the subspace can be taken to a `standard
configuration' $C_o$ in which all $1$'s and all $0$'s are clustered
together. 
Since the lattice is periodic, the $11$ and $00$ pairs can be
shifted to any other configuration $C'$ from $C_o$. Thus any configuration
$C$ can be taken to any configuration $C'$ via $C_o$.  

In steady state the right hand side of (\ref{master}) must vanish. A sufficient
condition for this is that
fluxes balance in pairs, i.e. for every flux out of a configuration $C$ to
a configuration $C^{'}$, there should be an incoming flux from another uniquely
determined configuration $C^{''}$ into $C$. This is the condition of
pairwise balance \cite{pairwise}:
\beq
        W(C^{''} \brt C) P(C^{''}) = W(C \brt C^{'}) P(C)
\label{pairwise}
\eeq
which is a generalization of the well known condition of detailed balance
\cite{kampen}.

For our problem, $C^{''}$ may be constructed as follows.
Let us denote the configuration $C$ symbolically as $1^{m_1} 0^{m_2}
1^{m_3} 0^{m_4} 1^{m_5} \cdots 0^{m_k}$, where there is a cluster of
$1$'s of size $m_1$, followed by $0$'s of length $m_2$, and so on, 
with a total of
$k$ such clusters. Consider a transition to a configuration
$C^{'} \equiv 1^{m_1} 0^{m_2-1} 1^2 0 1^{m_3 - 2} 0^{m_4} 1^{m_5}
\cdots 0^{m_k}$ (a pair of $11$'s jump to the left, i.e. move ($b$)).
One can always find a unique configuration e.g. $C^{''} \equiv
1^{m_1} 0^{m_2} 1^{m_3 - 2} 0 1^2 0^{m_4 - 1} 1^{m_5} \cdots 0^{m_k}$
which gives rise to $C$ (via move ($b$)). If the
outgoing transition involves a rearrangement at the left edge of a
cluster, the incoming transition involves a rearrangement at the right
edge of the same cluster.  Such an identification is possible also for
transitions involving moves ($a$), ($\bar{a}$) and ($\bar{b}$),
and ensures that $W(C^{''} \brt C) = W(C \brt C^{'})$. 
Thus  Eq.(\ref{pairwise}) is satisfied provided the
steady state probabilities obey
\beq
P(C) = P(C^{''}) = constant.
\label{equalprob}
\eeq
This means that in steady state, every allowed configuration
is equally likely.   
The constant appearing in Eq. (\ref{equalprob}) can be found on using the
normalization condition $\sum P(C)=1$. If $N_1 \equiv \rho^o_1 L$
and $N_2 \equiv \rho^o_2 L$ are the number of particles on the two 
sublattices, the total number of configurations in sector $(\rho^o_1,
\rho_2^o)$ is
${\cal {N}}={(^{L}C_{N_1})(^{L}C_{N_2})}$, where $^NC_M$ is the 
number of ways of choosing $M$ out of a total of $N$ objects, 
and hence $P(C) = 1/{\cal {N}}$.

In the thermodynamic limit $L$, $N_1$, and $N_2$
$\brt \infty$, with $\rho_1^o$, $\rho_2^o$ held constant, $P(C)$ approaches the
product measure form
\beq
{\prod_{i}}p(\sigma_{i})p(\tau_{i+{1\over
2}}) = {\rho_1^o}^{N_1}(1-\rho_1^o)^{L - N_1} 
{\rho_2^o}^{N_2} (1 -\rho_2^o)^{L - N_2}
\label{product}
\eeq
This form of the steady state holds also for a higher-dimensional
generalization of the model involving
rules (a) and (b) and their reverses along
the sites and bonds in the $d$ directions of a simple cubic lattice. 

The product measure weight in the steady state implies that correlation functions
on different sites decouple. This then allows the current of $\sigma$ particles
\beq
J_1 =  (p_2 - p_1) \langle {(1+\sigma_i)\over 2} {(1 -
\sigma_{i+1})\over 2} {(1 - \tau_{i + {1 \over 2}})\over 2}\rangle 
+ (p_2 - p_1)
\langle {(1+\sigma_{i+1})\over 2} {(1 - \sigma_i)\over 2}
{(1+\tau_{i + {1 \over 2}})\over 2} \rangle 
\label{currA}
\eeq
to be found explicitly,
\beq
J_1= (p_2-p_1) \rho_1 (1 - \rho_1) (1 - 2 \rho_2)
\label{currAA}
\eeq
The  first term in Eq. (\ref{currA}) comes
from a particle hopping between sites $i$ and $i+1$ in the absence of a particle
at site $i+{1\over 2}$ while the second is for hopping in the
presence of a particle on the site in between. A similar expression
holds for the $\tau$-current $J_2$,
\beq
J_2 = (p_2 - p_1) \rho_2 (1 - \rho_2) (1 - 2 \rho_1).
\label{currB}
\eeq

The product measure form also allows us to find the roughness exponent
of an associated height model, where the height fields associated
with $\{\sigma_i\}$ and $\{\tau_i\}$ are respectively
$h_{1i} = \sum_{k=1}^{i} (\sigma_k - \langle \sigma_k \rangle)$ and 
$h_{2i} = \sum_{k=1}^{i} (\tau_k - \langle \tau_k \rangle)$.  
Fluctuations of the height field are characterized by
the root-mean-square height difference
$G_1(r) =\sqrt{\langle (h_{1i+r} - h_{1i})^2 \rangle}$, with $G_2(r)$
defined similarly in terms of $\{ h_{2i} \}$.
Using the fact that $\langle \sigma_0 \sigma_k \rangle = 
\langle \sigma_0 \rangle \langle \sigma_k\rangle $ for $k \neq 0$
and $1$ for $k = 0$, we find
$G_1(r)=G_2(r) \sim r^{1/2}$. Thus the roughness exponent $\chi$
defined by the growth of the root-mean-squared height fluctuations is
\beq
\chi = 1/2
\label{rough}
\eeq
for both height fields. Evidently, this value will also characterize
fluctuations of linear combinations of the height fields $h_1$ and $h_2$,
which arise when we deal with mode coordinate fields in the next section.

\figone{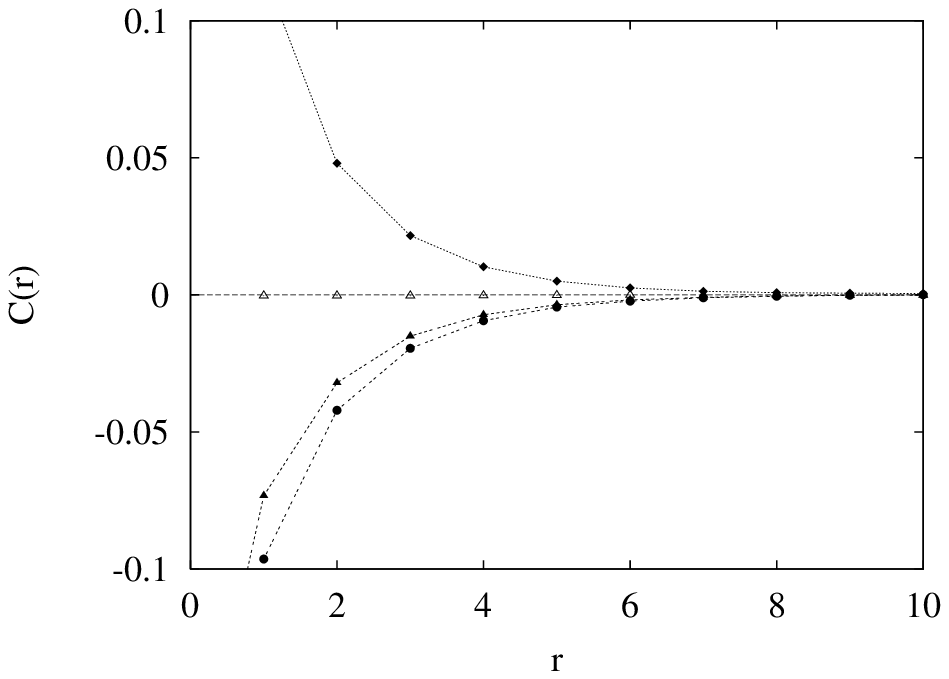}{The decay of correlation functions away from the
line of pairwise balance is shown for  $C_{11}$
(filled circle), $C_{22}$ (filled diamond) and $C_{12}$ (filled
triangle).  When pairwise balance does hold, the correlation function
vanishes (empty triangles).}{nonprod}

We close this section with numerical results for correlation functions
away from the pairwise balance (PB) locus. We set $r_2=p_1=0$, and investigate
what happens if we move away from the PB locus $r_1=p_2$.
We studied the spatial correlation
function $C(r) = \langle \sigma_i \sigma_{i+r} \rangle -
\langle \sigma_i \rangle \langle \sigma_{i+r}\rangle$
by Monte-Carlo simulation, and show our results for different values of $r_1$
and $p_2$
in Fig. \ref{nonprod}. As expected, for the PB case
$r_1 = p_2 = 1$ (empty triangles in Fig. \ref{nonprod}), the correlation
length is zero.  Away from PB, with $r_1 = 1/3$ and $p_2 = 1$, 
the three curves for $C_{11}$, $C_{22}$ and $C_{12}$ representing respectively
the intra-sublattice correlation functions for sublattices 1 and 2 and the
inter-sublattice correlation function all decay differently,
but with a finite correlation length.  (Fig. \ref{nonprod}).
This shows that although product measure does not hold away from the 
pairwise balance locus $r_1=p_2$, correlations are short-ranged
so that the behaviour on large length and time
scales is expected to be similar to that in the pairwise balance case.

{\subsection {Continuum equations and kinematic waves}}
\label{contrevisited} 

The expressions (\ref{currAA}) and (\ref{currB}) for the current help us
to construct approximate continuum
equations for the evolution of the density fluctuation
fields. The starting point is the pair of continuity equations
\beq
{{\partial \rho_{m}} \over {\partial t}} = - {{\partial} \over
{\partial x}}J_{m}(\rho_1,\rho_2)~~~~~~~~~~m=1,2
\label{continuity}
\eeq
where $\rho_1(x,t)$ and $\rho_2(x,t)$
are coarse-grained densities at a mesoscopic scale and $J_1(x,t)$ and $J_2(x,t)$
are the corresponding currents. Each of these currents is written as the sum of 
three parts,
\beq
J_{m} = J_{m}^{systematic} + J_{m}^{diffusive} + \eta_{m}.
\label{J}
\eeq                   
The systematic contributions $J_m^{systematic}$ 
at the mesoscopic scale are assumed to be given by the expressions
(\ref{currAA}) and (\ref{currB}) for the currents in an infinite
system. The diffusive part $J_m^{diffusive}$ arises
from local density inhomogeneities and is  taken to be
$ - D_{m}{{\partial \rho_{m}}\over{\partial x}}$.
Finally $\eta_{m}$ is a noise term added to mimic stochasticity
at the mesoscopic level; we consider uncorrelated white noise with
$\langle \eta_m \rangle = 0$ and $\langle\eta_m(x,t)\eta_m(x',t')\rangle=\Gamma
\delta(x-x')\delta(t-t')$. These continuum equations have the same
symmetries as the lattice model and hence would be expected to exhibit
the same behaviour on large length and time scales.   

Writing $\rho_1 = \rho_1^o + \tilde{\rho_1}$, and $\rho_2 = \rho_2^o +
\tilde{\rho_2}$ (where $\rho^o$'s are fixed average densities and
$\tilde{\rho}$'s are fluctuations) and using Eqs. (\ref{currA}), (\ref{currB}),
\ref{continuity} and \ref{J}, one can write down coupled equations
governing the evolution of $\tilde\rho$'s. We write these in
terms of the height functions $h_1 = \int^x \tilde\rho_1(x^{'},t) dx^{'}$ and
$h_2 = \int^x \tilde\rho_2(x^{'},t) dx^{'}$,
the continuum analogs of the discrete functions
$h_{1i}$ and $h_{2i}$ defined in the previous section. We find 
\begin{eqnarray}
{{\partial h_1} \over {\partial t}} &=&
- r^{'} (1 - 2\rho_1^o)(1 - 2\rho_2^o){{\partial h_1} \over {\partial x}}
+ 2 r^{'} {\rho_1^o}(1 - \rho_1^o){{\partial h_2} \over {\partial x}}
+ D_1 {{\partial^2 h_1} \over {\partial x^2}} \nonumber \\
&+& 2 r^{'} (1 - 2\rho_1^o){{\partial h_1} \over {\partial x}}{{\partial h_2}
\over {\partial x}}
+ r^{'} (1 - 2\rho_2^o)({{\partial h_1} \over {\partial x}})^2
- 2r^{'} ({{\partial h_1} \over {\partial x}})^2
({{\partial h_2} \over {\partial x}}) + \eta_1(x,t)
\label{h1}
\end{eqnarray}
and
\begin{eqnarray}
{{\partial h_2} \over {\partial t}} &=&
- r^{'}(1 - 2\rho_1^o)(1 - 2\rho_2^o){{\partial h_2} \over {\partial x}}
+ 2 r^{'}{\rho_2^o}(1 - \rho_2^o){{\partial h_1} \over {\partial x}} +
D_2 {{\partial^2 h_2} \over {\partial x^2}} \nonumber \\
&+& 2r^{'} (1 - 2\rho_2^o){{\partial h_1} \over {\partial x}}{{\partial h_2}
\over {\partial x}}
+ r^{'}(1 - 2\rho_1^o)({{\partial h_2} \over {\partial x}})^2
- 2r^{'} ({{\partial h_2} \over {\partial x}})^2
({{\partial h_1} \over {\partial x}}) + \eta_2(x,t)
\label{h2}
\end{eqnarray}   
where $r'=(r_1-r_2)$.

Let us define $a = r^{'}(1 - 2\rho_1^o)(1 - 2\rho_2^o)$,
$b = r^{'}{\rho_1^o}(1 - \rho_1^o)$, $c = r^{'}{\rho_2^o}(1 -
\rho_2^o)$, $\kappa_1 = r^{'}(1 - 2\rho_1^o)$, and $\kappa_2 = r^{'}(1 -
2\rho_2^o)$. It is apparent that by taking linear combinations one can
construct eigenmode fields $h_{\pm} = \sqrt{c} h_1 \pm \sqrt{b} h_2$
which decouple at the linear level. These fields describe wave-like modes
\cite{lighthill} travelling with speeds $c_{\pm} = - a \pm
2\sqrt{bc}$. The time evolutions of these fields $h_{\pm}$ are governed
by 
\begin{eqnarray}
{{\partial h_+} \over {\partial t}} &=&
c_+{{\partial h_+} \over {\partial x}}
+ D {{\partial^2 h_+} \over {\partial x^2}}
+ {3\over 2}({\kappa_2 \over \sqrt{c}} + {\kappa_1 \over \sqrt{b}})
({{\partial h_+} \over {\partial x}})^2
+ ({\kappa_2 \over \sqrt{c}} - {\kappa_1 \over \sqrt{b}})
{{\partial h_+} \over {\partial x}}{{\partial h_-} \over {\partial x}}
\nonumber \\
&-& {1\over 2}({\kappa_2 \over \sqrt{c}} + {\kappa_1 \over \sqrt{b}})
({{\partial h_-} \over {\partial x}})^2
-{1\over {2\sqrt{bc}}}({{\partial h_+} \over {\partial x}})
\left[({{\partial h_+} \over {\partial x}})^2
-({{\partial h_-} \over {\partial x}})^2\right] + \eta_+(x,t);
\nonumber \\            
{{\partial h_-} \over {\partial t}} &=&
c_{-}{{\partial h_-} \over {\partial x}}
+ D {{\partial^2 h_-} \over {\partial x^2}}
+ {3\over 2}({\kappa_2 \over \sqrt{c}} - {\kappa_1 \over \sqrt{b}})
({{\partial h_-} \over {\partial x}})^2
+ ({\kappa_2 \over \sqrt{c}} + {\kappa_1 \over \sqrt{b}})
{{\partial h_+} \over {\partial x}}{{\partial h_-} \over {\partial x}}
\nonumber \\
&-& {1\over 2}({\kappa_2 \over \sqrt{c}} - {\kappa_1 \over \sqrt{b}})
({{\partial h_+} \over {\partial x}})^2
-{1\over {2\sqrt{bc}}}({{\partial h_-} \over {\partial x}})
\left[({{\partial h_+} \over {\partial x}})^2
-({{\partial h_-} \over {\partial x}})^2\right] + \eta_-(x,t).
\label{h+-}
\end{eqnarray} 
The new noise terms $\eta_{\pm} = \sqrt{c} \eta_1 \pm \sqrt{b}
\eta_2$, are also delta-correlated. We have assumed $D_1
= D_2 = D$, though this may not be preserved in the effective long
wavelength equations.  The fields $h_+$ and $h_-$ are coupled 
at the nonlinear level, so that each wave
influences the dissipation of fluctuations of the other.
We consider the dissipation properties of the waves in
the next section for different sublattice filling fractions $\rho_{1}^{o}$ and
$\rho_{2}^{o}$ of the sublattices. 

\noindent {\section {\bf Dissipation of the waves and dynamical exponents}}
\label{dynexplat}
If it happens that some of the coefficients of (\ref{h+-}) vanish 
for certain choices of densities $\rho_{1}^{o}$ and $\rho_{2}^{o}$, the
evolution equations have special symmetries and this can have important 
implications for the long-time dynamics. As discussed below, there are
three different symmetries which arise in the coupled field problem,
each corresponding to a different set of dynamical exponents.
The dynamical exponents associated with the
wave modes, may differ. We have considered three special pairs of
densities $(\rho^{o}_{1}, \rho^{o}_{2})$ corresponding to three different
symmetries.  

\noindent {\subsection {\bf Symmetries of the equations:}}
\label{symmetries} 
To facilitate subsequent discussions let us first consider the case of a
single field $h$, and list four different symmetries \cite{stanley} for its
evolution.

(a) $RI$ symmetry: Invariance under up-down Reflection ($R$) symmetry
$h \rt -h$ and under Inversion ($I$) of space $x \rt -x$.

(b) $R{\bar{I}}$ symmetry: Invariance under $h \rt -h$ and {\it
not} under $x \rt -x$.

(c) ${\bar{R}}I$ symmetry: Invariance under $x \rt -x$ and {\it not}
under $h \rt -h$.

(d) ${\bar{R}\bar{I}}$ symmetry: Invariance neither under $x \rt -x$,
nor under $h \rt -h$. 

Since  ${\partial h\over\partial t}$ is odd an equation of motion which
contains only terms odd in $h$ will be said to have $R$ inversion symmetry,
any term that is even in $h$ will be said to break $R$ symmtery. Accordingly:
A term like ${\partial^2 h} \over {\partial x^2}$ obeys $RI$
symmetry. Terms like $({{\partial h} \over {\partial x}})$ and
$({{\partial h} \over {\partial x}})^3$ obey $R{\bar{I}}$
symmetry.  The ${\bar{R}}I$ symmetry is respected by the term
$({{\partial h} \over {\partial x}})^2$, while a term like
$({{\partial h} \over {\partial x}})$ added to it
breaks that and gives rise to ${\bar{R}}{\bar{I}}$ symmetry.

To illustrate the occurrence of different types of symmetries in our
coupled-field problem, we consider three special
pairs of densities ($\rho_1^o$,$\rho_2^o$).                 

(I) For $\underline{\rho_1^o = \rho_2^o = {1 \over 2}}$,  (\ref{h+-})
reduces to a pair of coupled equations, with linear and first and
second derivative terms and cubic gradient nonlinearities:
\begin{eqnarray}
{{\partial h_+} \over {\partial t}} &=&
2\sqrt{bc}{{\partial h_+} \over {\partial x}}
+ \nu {{\partial^2 h_+} \over {\partial x^2}}
-{1\over {2\sqrt{bc}}}({{\partial h_+} \over {\partial x}})
\left[({{\partial h_+} \over {\partial x}})^2
-({{\partial h_-} \over {\partial x}})^2\right] + \eta_+(x,t);
\nonumber \\
{{\partial h_-} \over {\partial t}} &=&
-2\sqrt{bc}{{\partial h_-} \over {\partial x}}
+ \nu {{\partial^2 h_-} \over {\partial x^2}}
-{1\over {2\sqrt{bc}}}({{\partial h_-} \over {\partial x}})
\left[({{\partial h_+} \over {\partial x}})^2
-({{\partial h_-} \over {\partial x}})^2\right] + \eta_-(x,t).
\label{case1}
\end{eqnarray}
These equations describe two kinematic waves moving with speed $c_+=r^{'}/2$ and
$c_-=-r^{'}/2$. The nonlinear couplings imply that each wave influences the
evolution of the other. In order to study the dissipation of say
the $+$mode, it is essential to move to the frame which co-moves with it. This
is accomplished by a Galilean shift $x \rt x + c_+t$, and $t \rt t$.
In this frame, the evolution equations become:
\begin{eqnarray}
{{\partial h_+} \over {\partial t}} &=&
 \nu {{\partial^2 h_+} \over {\partial x^2}}
-{1\over {2\sqrt{bc}}}({{\partial h_+} \over {\partial x}})
\left[({{\partial h_+} \over {\partial x}})^2
-({{\partial h_-} \over {\partial x}})^2\right] + \eta_+(x,t);
\nonumber \\
{{\partial h_-} \over {\partial t}} &=&
-4\sqrt{bc}{{\partial h_-} \over {\partial x}}
+ \nu {{\partial^2 h_-} \over {\partial x^2}}
-{1\over {2\sqrt{bc}}}({{\partial h_-} \over {\partial x}})
\left[({{\partial h_+} \over {\partial x}})^2
-({{\partial h_-} \over {\partial x}})^2\right] + \eta_-(x,t).
\label{case1+}
\end{eqnarray}
Evidently, in this frame, the $-$mode has a speed $c_{-} - c_{+} =
-4\sqrt{bc}$. The pair of Eqs. (\ref{case1+}) are invariant under
$h_+ \rt -h_+$, $h_- \rt -h_-$ but not $x \rt -x$, because of the
linear ${{\partial h_-} \over {\partial x}}$ and cubic nonlinear terms.
The $R{\bar{I}}$ symmetry holds in the rest frame of $h_+$ mode.

Similarly the dissipation of the $-$mode can be
studied by going to a frame which co-moves with the $-$mode.
It is easily seen that $R{\bar I}$ symmetry holds in this frame as well. 

Let us recall what happens when $R{\bar I}$ symmetry holds in the
case of a single field $h$.
In the Edwards-Wilkinson (EW) equation \cite{EW}, the 
presence of an additional cubic term like
$({{\partial h} \over {\partial x}})^3$ reduces the $RI$ symmetry
to $R{\bar{I}}$ symmetry. About the linear fixed point (with $\chi = 1/2$),
the cubic term has the
same na\"ive scaling dimension as $({{\partial^2 h} \over {\partial x}^2})$
--- both scale as $b^{-{3/2}}$ when $x \rt bx$ and $h \rt b^{\chi}h$.
Such a marginal cubic term is known
\cite{spohn1,spohn2,spohn3,spohn4} to introduce
logarithmic factors in the behaviour of the height-height correlation
functions. Using mode-coupling \cite{spohn1} and
dynamical renormalization group \cite{spohn2} calculations, it was found
that the correlation function
$F(t) = \sqrt{\langle [h(x,t) - h(x,0)]^2 \rangle}$
grows as $t^{1/4}(log(t))^{1/8}$. Recalling that $F(t) \sim t^{\beta}$
with $\beta=\chi/z$, we see that despite the lack of $I$ symmetry,
the critical exponents $\beta$ and $z$ do not change from their EW values
$1/4$ and $2$, respectively. 

In our case with two coupled fields,
the cubic gradient terms again have the same na\"ive scaling
dimension as the linear second order term, but they are more complicated
than just $({{\partial h}\over{\partial x}})^3$. We might guess nevertheless
due to the symmetry that each of the $+$ and $-$modes have $z = 2$.
We will present numerical evidence in section 5.2, which confirms this
and shows that there are similar multiplicative logarithmic factors. 

(II) For \underline{${\rho_1^{o} = 1/2}$ and ${\rho_2^{o} \neq 1/2}$},
the equations for the mode fields reduce to:
\begin{eqnarray}
{{\partial h_+} \over {\partial t}} &=&
2\sqrt{bc}{{\partial h_+} \over {\partial x}}
+ \nu {{\partial^2 h_+} \over {\partial x^2}}
+ {3\over 2}{\kappa_2 \over \sqrt{c}} ({{\partial h_+} \over {\partial x}})^2
+ {\kappa_2 \over \sqrt{c}}
{{\partial h_+} \over {\partial x}}{{\partial h_-} \over {\partial x}}
\nonumber \\
&-& {1\over 2}{\kappa_2 \over \sqrt{c}}({{\partial h_-} \over {\partial x}})^2
-{1\over {2\sqrt{bc}}}({{\partial h_+} \over {\partial x}})
\left[({{\partial h_+} \over {\partial x}})^2
-({{\partial h_-} \over {\partial x}})^2\right] + \eta_+(x,t);
\nonumber \\
{{\partial h_-} \over {\partial t}} &=&
-2\sqrt{bc}{{\partial h_-} \over {\partial x}}
+ \nu {{\partial^2 h_-} \over {\partial x^2}}
+ {3\over 2}{\kappa_2 \over \sqrt{c}}
({{\partial h_-} \over {\partial x}})^2
+ {\kappa_2 \over \sqrt{c}}
{{\partial h_+} \over {\partial x}}{{\partial h_-} \over {\partial x}}
\nonumber \\
&-& {1\over 2}{\kappa_2 \over \sqrt{c}}
({{\partial h_+} \over {\partial x}})^2
-{1\over {2\sqrt{bc}}}({{\partial h_-} \over {\partial x}})
\left[({{\partial h_+} \over {\partial x}})^2
-({{\partial h_-} \over {\partial x}})^2\right] + \eta_-(x,t).
\label{case2}
\end{eqnarray} 
Going to either of the frames in which the $+$mode or the $-$ mode are at rest,
we see that with the cubic nonlinearities the ${\bar{R}}{\bar{I}}$ symmetry
applies for each of the fields; $R$ symmetry is broken by
by quadratic nonlinear terms, and $I$ is broken because of linear first
order and cubic terms. The most relevant terms at the linear fixed point
are the quadratic nonlinear terms. Thus we would expect that these terms
would govern the dissipation and give rise to the Kardar-Parisi-Zhang (KPZ)
value $z = 3/2$ \cite{KPZ} for both the modes.

(III) For $\underline{\rho_1^{o} = \rho_2^{o} \neq 1/2}$, we have
$\kappa_1 = \kappa_2$ and $c = b$, and the following pair of equations
hold:
\begin{eqnarray}
{{\partial h_+} \over {\partial t}} &=&
c_+{{\partial h_+} \over {\partial x}}
+ D {{\partial^2 h_+} \over {\partial x^2}}
+ {3}{\kappa_2 \over \sqrt{c}}
({{\partial h_+} \over {\partial x}})^2
\nonumber \\
&-& {\kappa_2 \over \sqrt{c}}
({{\partial h_-} \over {\partial x}})^2
-{1\over {2\sqrt{bc}}}({{\partial h_+} \over {\partial x}})
\left[({{\partial h_+} \over {\partial x}})^2
-({{\partial h_-} \over {\partial x}})^2\right] + \eta_+(x,t);
\nonumber \\
{{\partial h_-} \over {\partial t}} &=&
c_{-}{{\partial h_-} \over {\partial x}}
+ D {{\partial^2 h_-} \over {\partial x^2}}
+ 2{\kappa_2 \over \sqrt{c}}
{{\partial h_+} \over {\partial x}}{{\partial h_-} \over {\partial x}}
\nonumber \\
&~&-{1\over {2\sqrt{bc}}}({{\partial h_-} \over {\partial x}})
\left[({{\partial h_+} \over {\partial x}})^2
-({{\partial h_-} \over {\partial x}})^2\right] + \eta_-(x,t).
\label{case3}
\end{eqnarray} 
Here, an interesting situation
arises. In the co-moving frame of the $-$mode, the pair of equations
are invariant under $h_- \rt -h_-$ and $h_+ \rt h_+$ but not under
$x \rt -x$. Thus the $h_-$ field has $R{\bar{I}}$ symmetry, while the
moving $h_+$ field has ${\bar R}{\bar I}$ symmetry. The equations are invariant
under $h_-\to -h_-$ in any frame. The invariance under $x\to -x$ is broken (a)
by the linear ${\partial h_{\pm}\over\partial x}$ and (b) by the trilinear
terms. The effect of (a) can be shifted away by comoving; the (b) terms are
expected to provide logarithmic multiplicative corrections.
The same symmetries hold in the rest frame of the $+$ wave.
Based on these observations, we expect $z=2$ for the $-$mode
(perhaps with multiplicative logarithmic corrections), and $z = 3/2$ for the
$+$mode. This is precisely the weak dynamical scaling discussed in the
Introduction.

The quadratic nonlinear terms in Eq. (\ref{case3}) are exactly like
those obtained in \cite{ertas1,kardar}. Similarly the nonlinearities
in (\ref{case2}) are of the form obtained in \cite{bara1}. 
The crucial difference is that we have additional linear gradient couplings,
which bring in the
possibility of observing weak dynamical scaling in our coupled-field
system.  We now turn to the next step of checking numerically and
analytically our symmetry-based expectations.

\noindent {\subsection {\bf Growth exponents from Monte-Carlo
simulation:}}
\label{expfrommcsim}
For numerical simulation, we used a definition of height ${\tilde
h}_i(t)$ which differs slightly from $h_i(t)$ discussed so
far. Instead of defining heights as density integrated over space with
respect to a fixed site, we define ${\tilde h}_{1i}$ and ${\tilde
h}_{2i}$ as integrated densities but with respect to the first
particle, which is itself moving.  Such a definition was used earlier
in \cite{goutam}, and was found to markedly reduce the fluctuation in
the height-height correlation function. If the particles on a
particular sublattice are labelled $1,\cdots,N_P$, then the height
${\tilde h}_i(t) = n_i(t) - \langle n_i(t)\rangle$ where $n_i(t)$ is
the tag number of the particle at site $i$ and the subtracted part has
a linear time dependence.  If the site $i$ is empty, ${\tilde h}_i$ is
determined from the tags of the closest particles on either side by a
lever rule.  If $i_{o}(t)$  (whose
average value $= vt$ with $v = $ particle speed, is the location of the
particle $1$) one has ${\tilde h}_i(t) =
\sum_{k=i_o(t)}^{i}{\tilde{\rho}}_k - \rho^o i_o(t)$, using the fact that
$n_i(t) = \sum_{k=i_o(t)}^{i}({\tilde{\rho}}_k+\rho^o)$. In our
problem, the height-height  
correlation $\langle {\tilde{h}_{iA}(t)} - \tilde{h}_{iA}(0) \rangle$ grows
as $-v_{A}\rho^{o}_{1} t$, where $v_{A} = (1-\rho^{o}_1)(1-2\rho^{o}_2)$.
The corresponding continuum equation for $\tilde{h}_{1}(x,t)$ has
two more additional terms compared to (\ref{h1}), i.e. a constant
$-v_{A}\rho^{o}_{1}$ which can be removed by an appropriate shift,
and a noise term $\zeta = v_{A}t\tilde{\rho}_1(x,t)$.
The scaling dimension of $\zeta$ is lower ($\zeta \rt b^{-z} \zeta$) than
the noise $\eta_1$ ($\eta_1 \rt b^{-z/2-1/2}\eta_1$), and hence is less
relevant. 

We are interested in knowing the dissipation property of
the $\pm$ wave modes, so we numerically compute the height-height
correlation function of the variable $\tilde{h}_{\pm}$ in the co-moving
frame of the relevant mode, i.e. by a Galilean shift with speed $c_{\pm}$.
The new fields $\tilde{h}_{\pm}$ are the counterparts of $h_{\pm}$
defined earlier. We monitor the correlation functions 
\beq
S^{2}_{\pm}(t) = \langle \left[ {\tilde{h}}_{\pm}(x-c_{\pm}t,t)
- \tilde{h}_{\pm}(x,0)
+ J^{o}_{\pm}t + \rho^{o}_{\pm}c_{\pm}t \right]^2 \rangle
\label{corr+-}
\eeq
where the average $\langle \cdots \rangle$ is both over space and the
ensemble of configurations in the steady state, while
$J^{o}_{\pm} = \sqrt{c}v_{A}\rho^{o}_1 \pm \sqrt{b}v_{B}\rho^{o}_2$ and
$\rho^{o}_{\pm} = \sqrt{c}\rho^{o}_1 \pm \sqrt{b}\rho^{o}_2$.
We expect $S^2_{\pm}(t)$ to grow as $t^{\beta}$ with $\beta=\chi/z$.

We used lattice sizes $L= 54000$, while the number of history averages for each
case is mentioned in the figure captions. There are two curves in each of
the figures \ref{corr0.5_0.5},
\ref{corr0.5_0.33}, and \ref{corr0.33_0.33} below: for example in Fig.
\ref{corr0.5_0.5}(b) the
lower one represents $S^{2}_+$, and the upper one represents the height-height
correlation function of the moving $-$mode in the rest frame of the
$+$wave,  namely
$\hat{S}^{2}_- = \langle \left[ {\tilde{h}}_{-}(x-c_{+}t,t) -
\tilde{h}_{-}(x,0)
+J^{o}_{-}t + \rho^{o}_{-}c_{+}t \right]^2 \rangle$. 
Similarly $\hat{S}^{2}_+$ in Fig. \ref{corr0.5_0.5}(a)
is defined in the rest frame of the $-$mode.
The correlation functions $\hat{S}^2$'s
increase linearly with time as they sense the effect of a
moving kinematic wave.
In a discrete lattice simulation, $x$ in Eq. (\ref{corr+-})
gets replaced by discrete integers $i$. The kinematic wave speeds are
nevertheless real fractions, so height-height correlation functions
$S^{2}$ and ${\hat{S}}^{2}$ have oscillations of the period $1/c_{\pm}$;
these oscillations are noticeable for low $t$, but their relative contribution
dies down for larger $t$.

(I) We first discuss the height-height correlations for
$\rho_1^o = 0.5$ and $\rho_2^o = 0.5$ (Fig. \ref{corr0.5_0.5})
corresponding to ${\rm R}\bar {\rm I}$ symmetry. 
\figfour{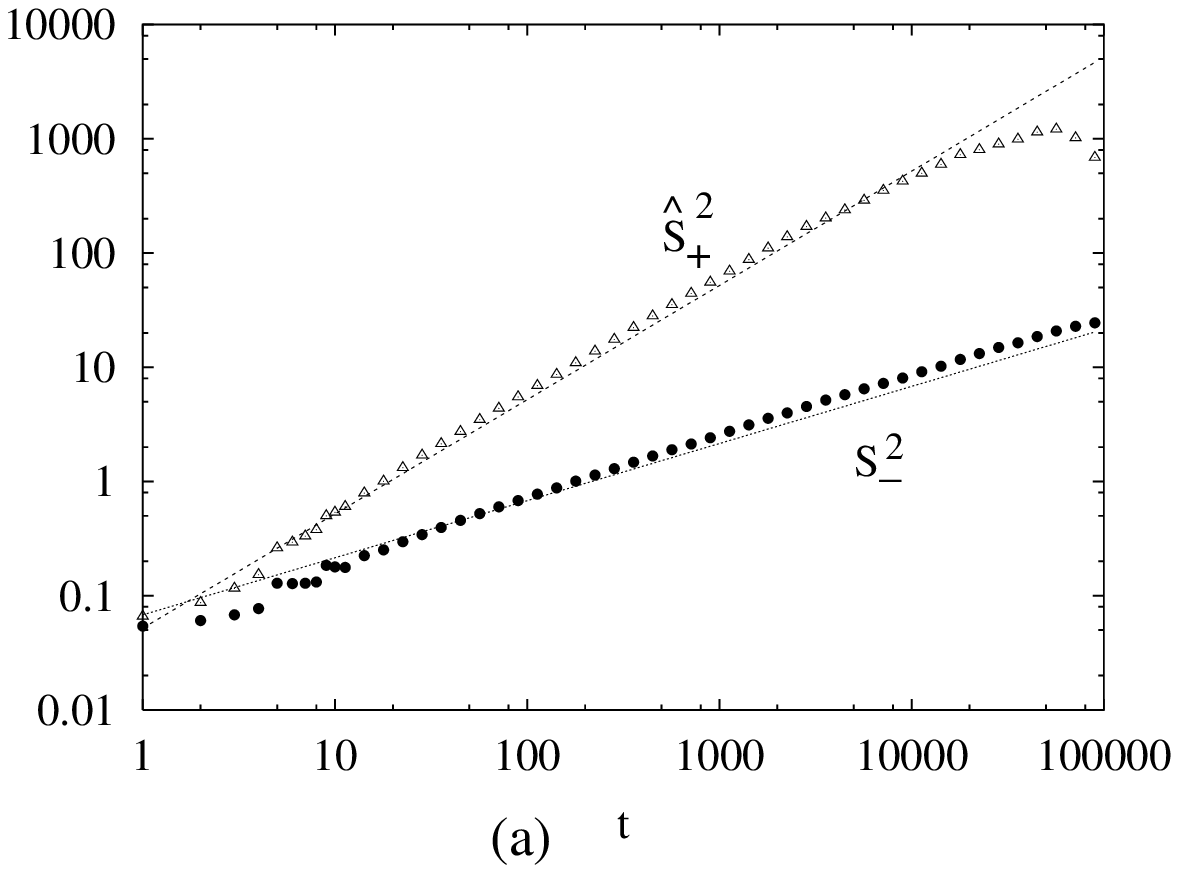}{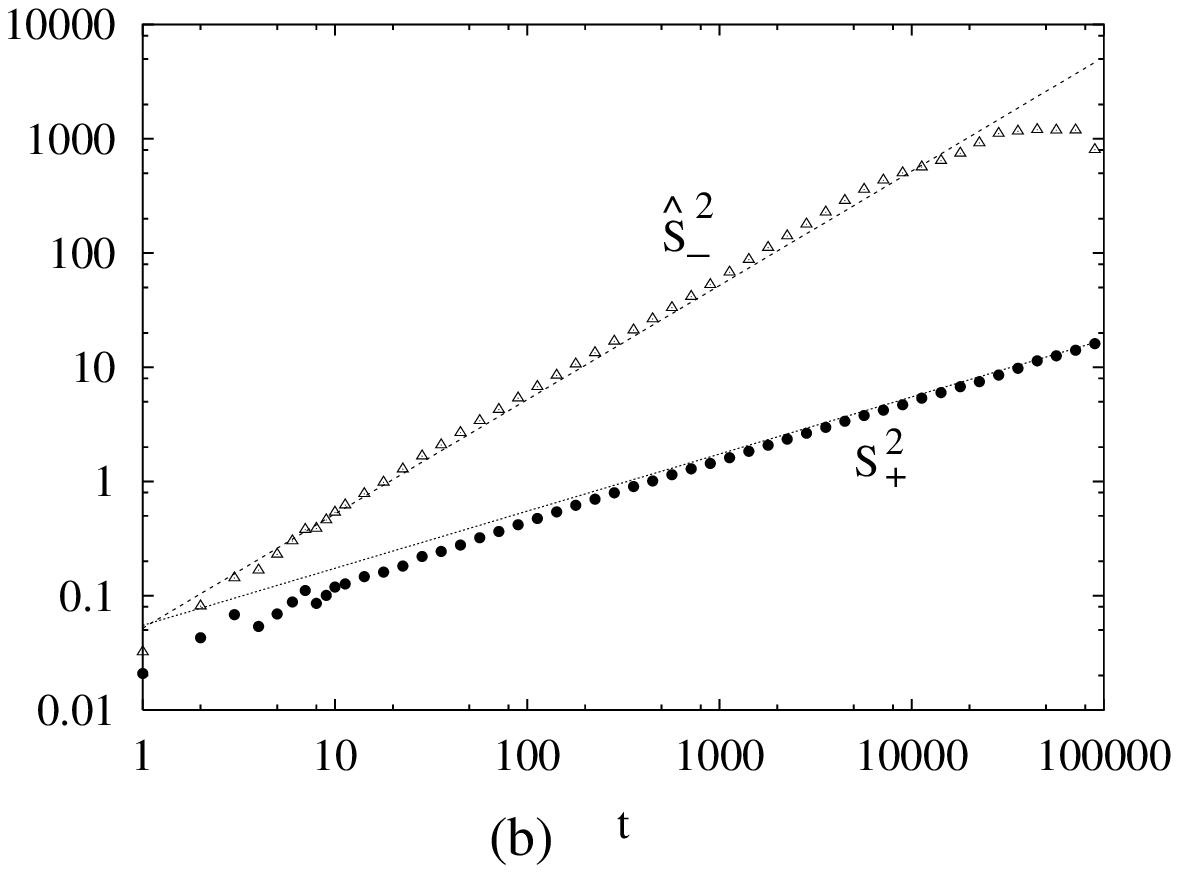}{Height correlation
functions for $\rho_1^o = \rho^o_2 = 0.5$ (${\rm R}\bar {\rm I}$
symmetry) averaged over $6$ histories. (a) In the rest frame
of the $-$ mode, $S^{2}_-$ grows as $\sim t^{1/2}$ implying $\beta =1/4$, 
while $\hat{S}^{2}_+ \sim t$. (b) In the rest frame of the $+$ mode, $S^{2}_+$ 
grows as $\sim t^{1/2}$, while  $\hat{S}^{2}_- \sim t$.  }{corr0.5_0.5}

The kinematic waves go around the whole system with a  time period $ t_o
= |L/c_{\pm}|$. With $L = 54000$ and $c_{\pm} = {\pm} r'/2 = \pm 1/4$
we have $t_o = 216000$ as $r'=1/2$ in our simulations. 
The curve for the mode which is moving in the rest frame of the other
should dip to a minimum at time $t_o$, and that is
why it shows a flattening at times around $t = t_o/2 = 108000$ when
the moving wave has travelled halfway around.

Figure \ref{corr0.5_0.5} gives strong evidence for $\beta=1/4$ for
both modes in their respective rest frames, when $\rho^o_1=\rho_2^o=1/2$.
Thus $z=\chi/\beta$ is equal to 2 for both modes.
More careful examination of the data reveals that there are 
actually multiplicative
logarithmic corrections to the leading power law behaviour. The  data
is consistent with a growth $S(t)\sim t^{1/4}(log(t))^{1/8}$,
as may arise from cubic nonlinearities as discussed above. Evidence of
this is seen in 
\figfour{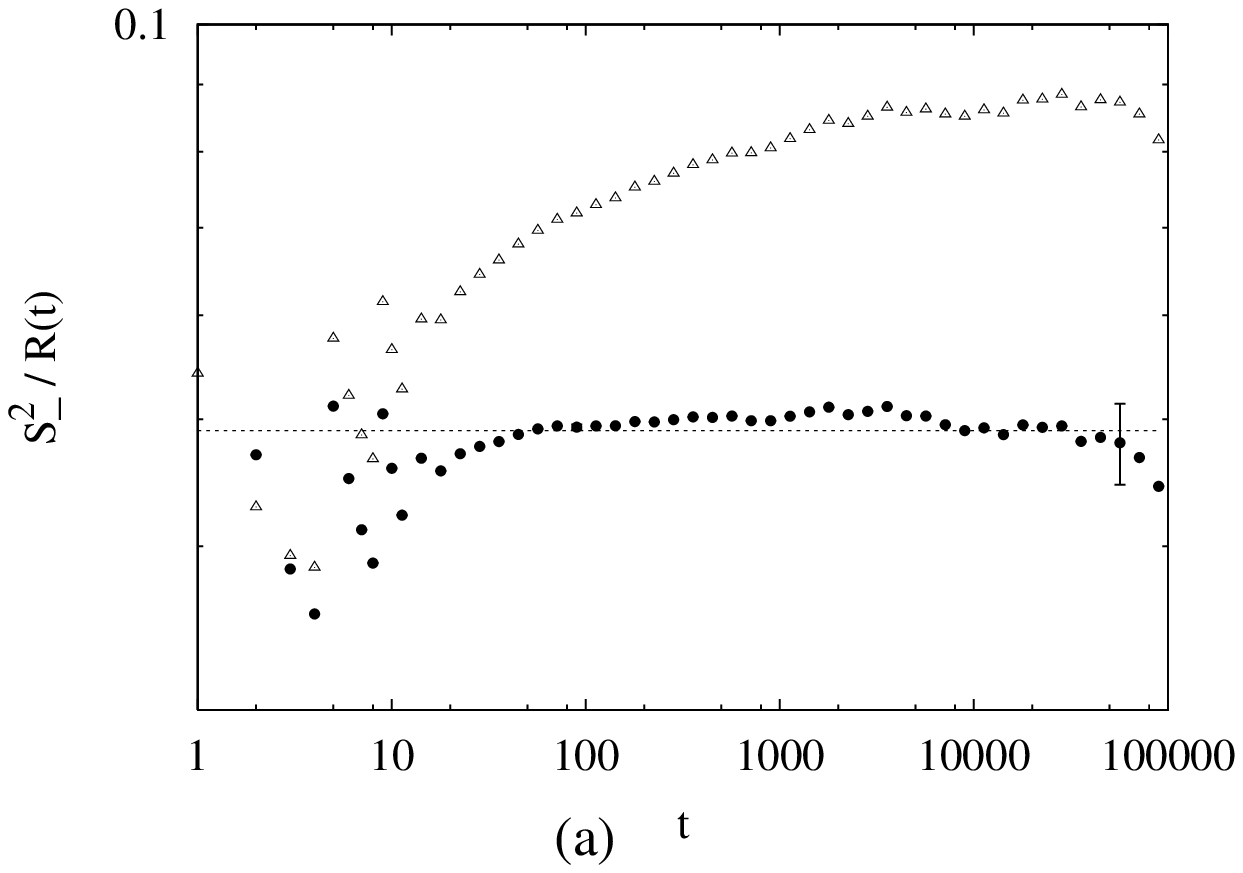}{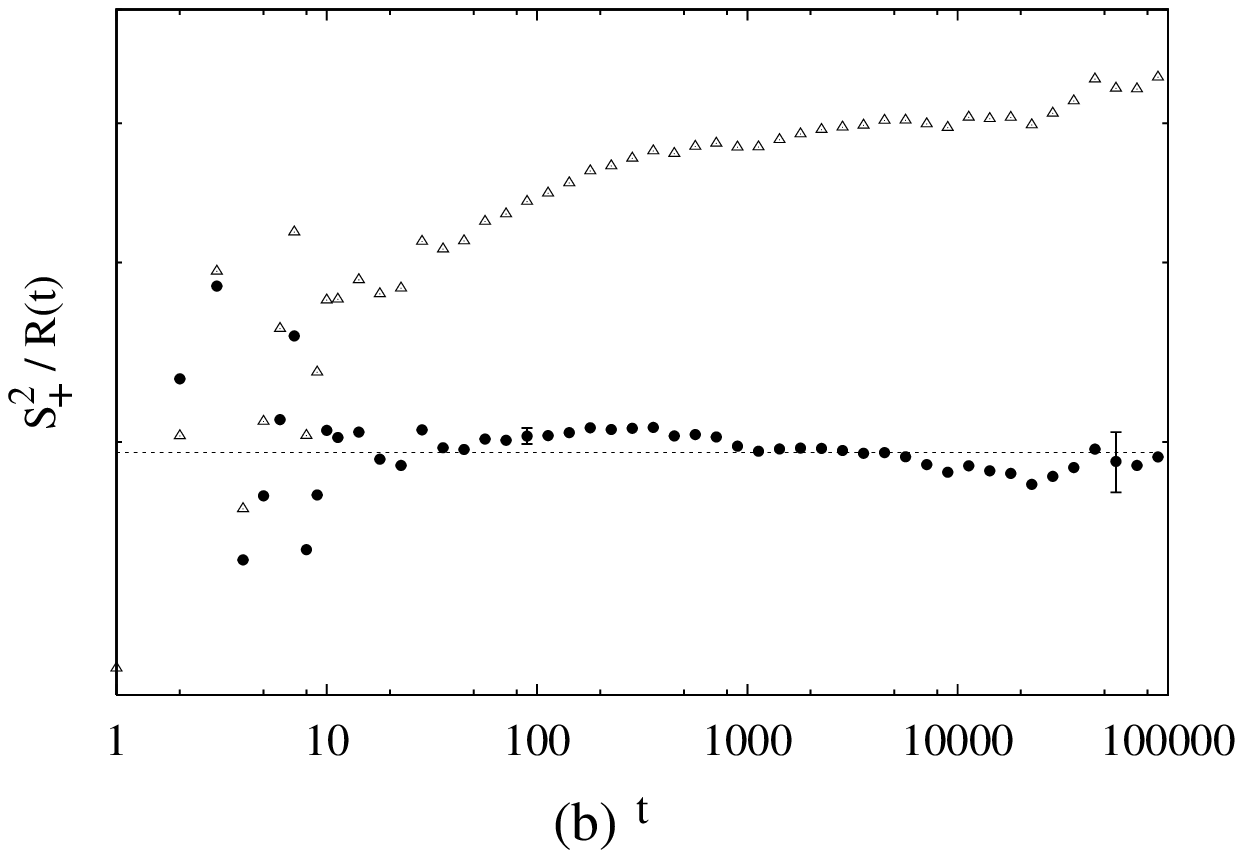}{Effect of scaling the rest-frame height 
correlation functions
of Fig. \ref{corr0.5_0.5} by a factor $R(t)= t^{1/2}$ (upper curves) 
and $R(t) = t^{1/2} (log(t))^{1/4}$ (lower curves). The flattening of the
latter confirms the presence of the multiplicative logarithms 
in both $S^2_{-}$ and $S^2_{+}$.}{log1}
Figs. \ref{log1}(a),(b) which indicate that the data
of Figs. \ref{corr0.5_0.5}(a),(b) have factors $(log(t))^{1/4}$
multiplying $t^{1/2}$. As discussed in section 5.1, the latter are
probably due to cubic nonlinearities.

(II) For $\rho_1^o = 0.5$ and $\rho_2^o = 1/3$, the system has ${\rm
R}\bar{\rm I}$ symmetry.  Figure \ref{corr0.5_0.33} shows that
both $S^{2}_{+}$ and $S_-^2$ grow as $\sim t^{2/3}$
 implying that $\beta = 1/3$ and
$z = 3/2$. Thus the dissipations of both the waves are KPZ-like. 

\figfour{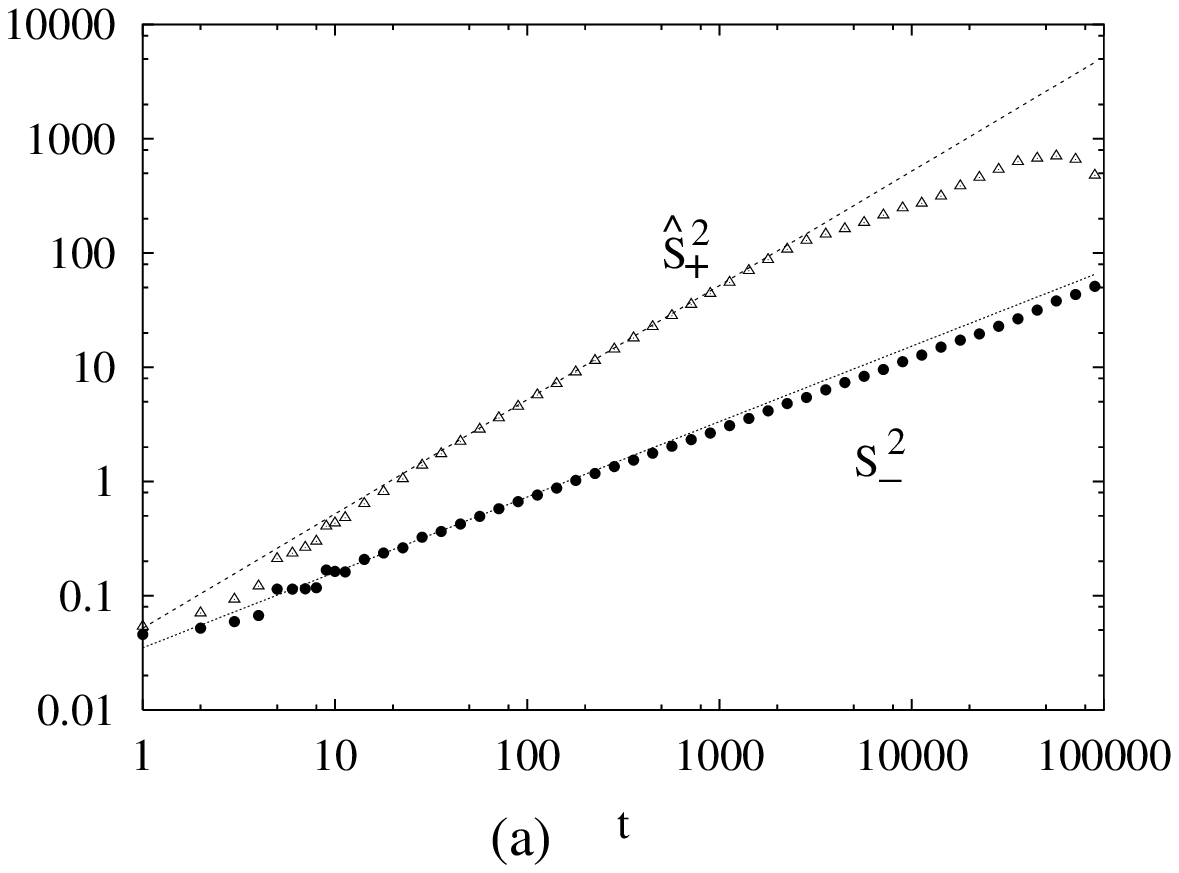}{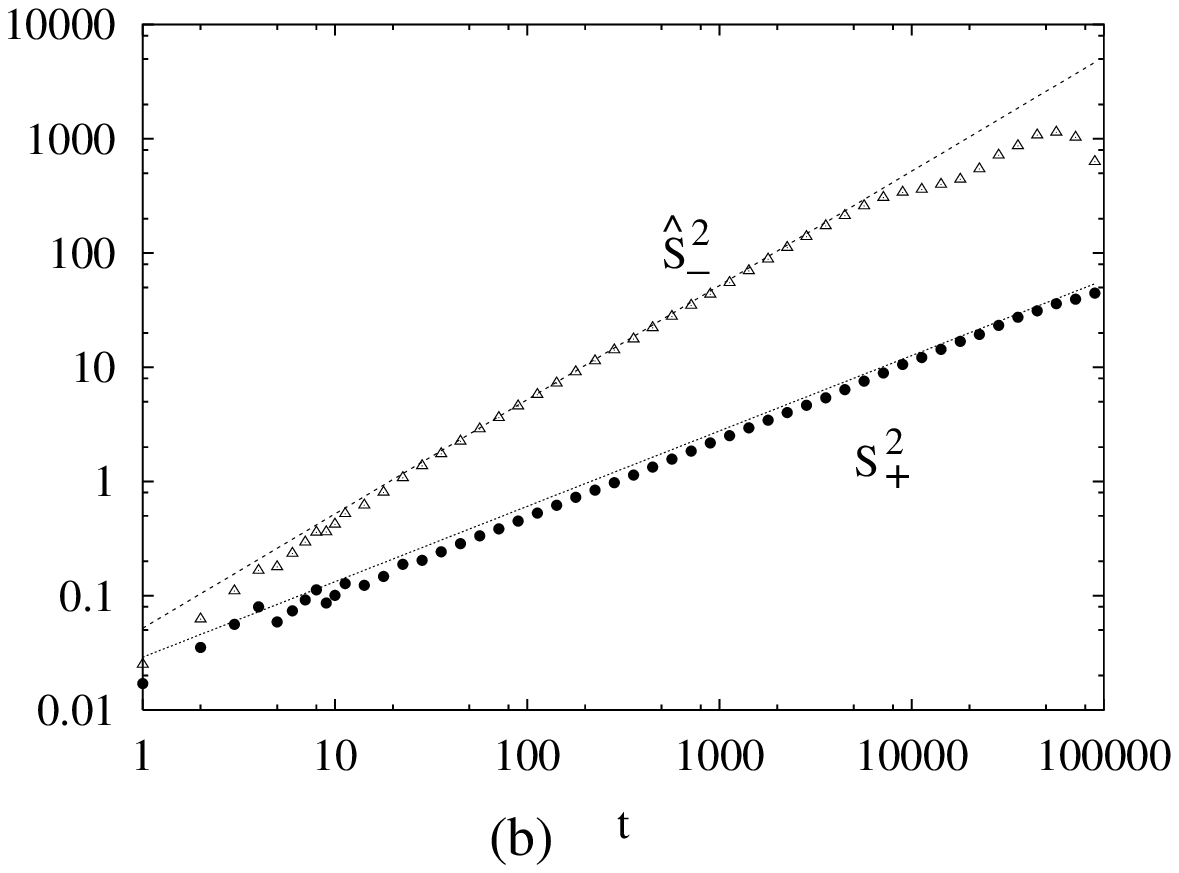}{Height correlation
functions for $\rho^o_1 = 1/2, \rho^o_2 = 1/3$ (${\rm R}\bar {\rm I}$
symmetry) averaged over 4 histories.   (a) $S^{2}_-$ grows
as $\sim t^{2/3}$ implying $\beta = 1/3$, while $\hat{S}^{2}_+ \sim t$. 
(b) $S^{2}_+$ too grows as
$\sim t^{2/3}$ and $\hat{S}^{2}_- \sim t$.}{corr0.5_0.33}
(III) Finally Fig. \ref{corr0.33_0.33} shows the height-height correlation
functions for $\rho_1^o = 1/3$ and $\rho_2^o = 1/3$ (${\rm R}\bar{\rm I}$ 
symmetry for one mode, $\bar {\rm R}\bar {\rm I}$ symmetry for the other).
\figfour{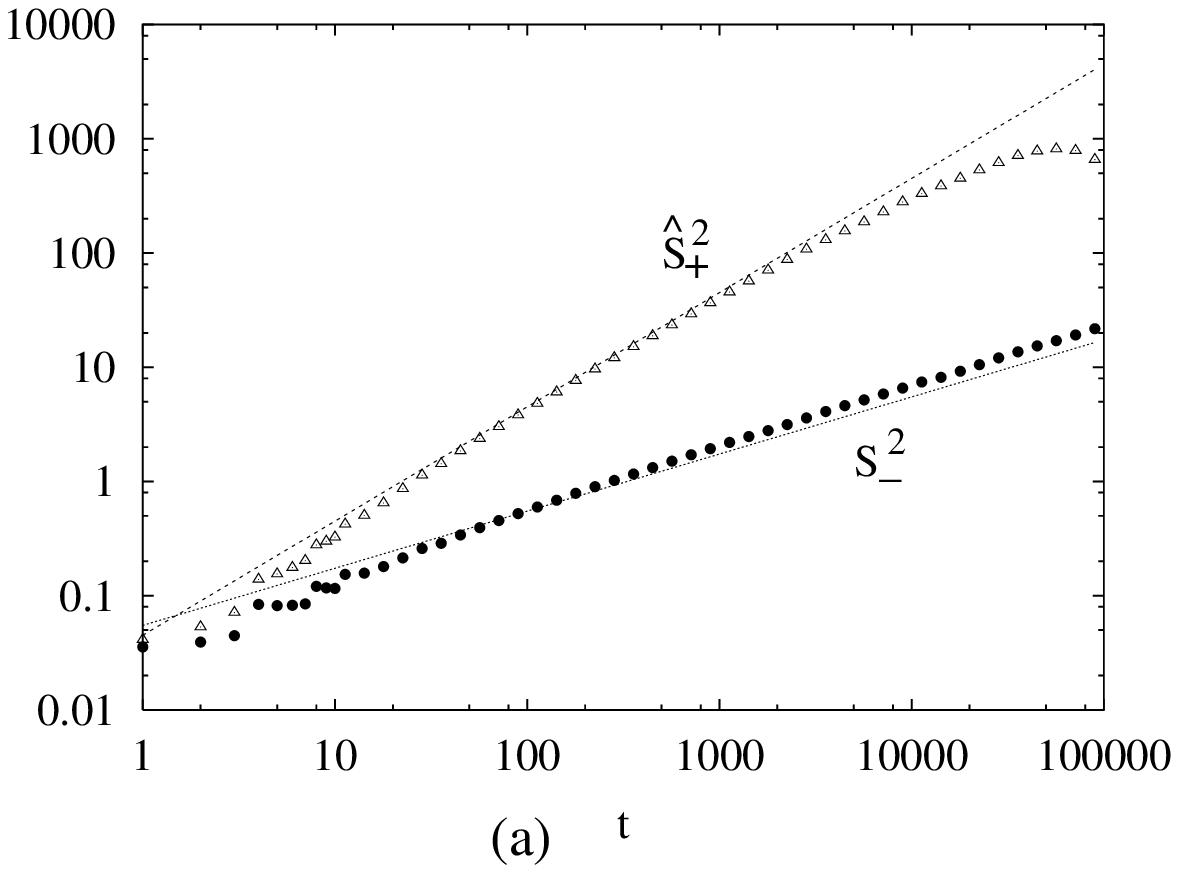}{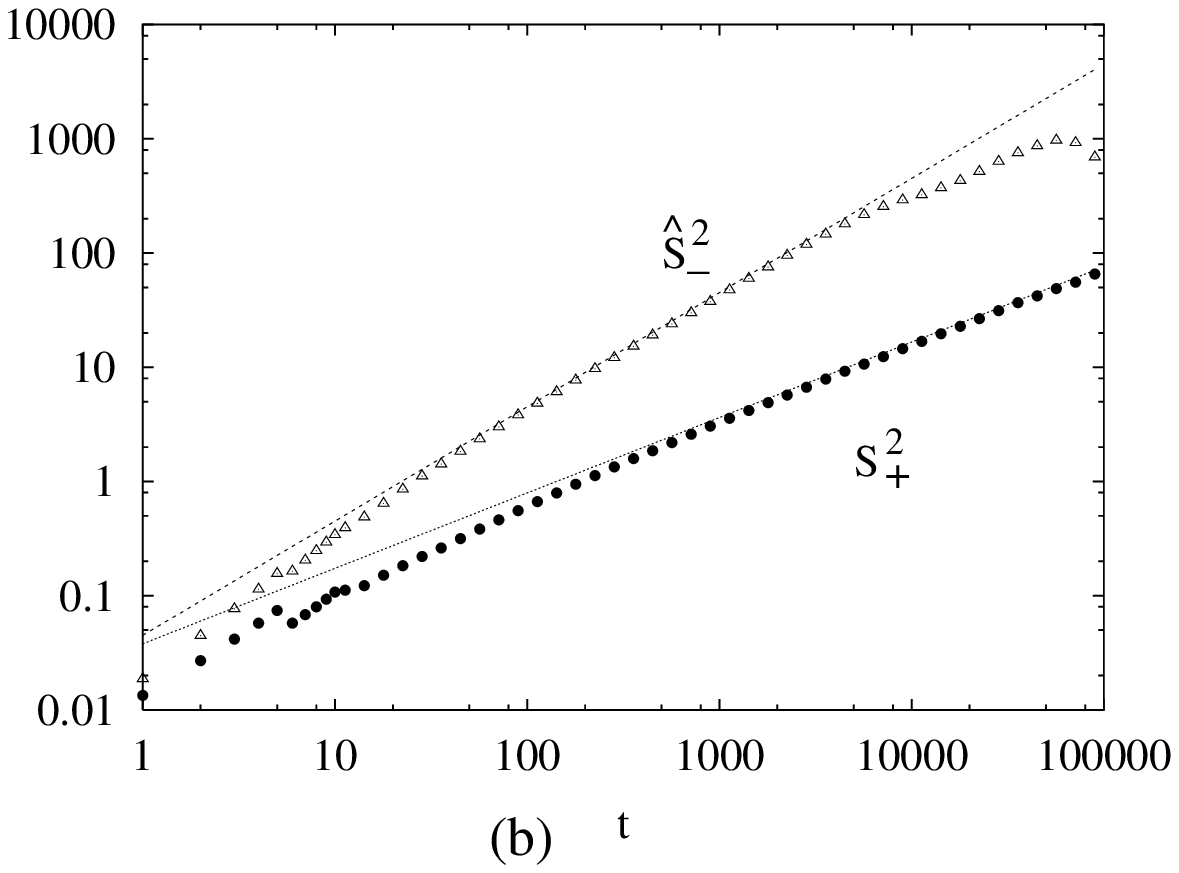}{ Height correlation
function for $\rho_1^o = \rho_2^o = 1/3$ (${\rm R}\bar{\rm I}$ symmetry
for one mode, $\bar {\rm R}\bar {\rm I}$ symmetry for the other) averaged over
12 histories.  (a) $S^{2}_-$ grows as $\sim t^{1/2}$ and $\hat{S}^{2}_+ \sim
t$. (b) $S^{2}_+$ grows as $\sim t^{2/3}$ and $\hat{S}^{2}_- \sim
t$. }{corr0.33_0.33} 
We see that $S^{2}_-$ grows as $\sim t^{1/2}$ with an indication of
multiplicative logarithmic corrections, indicating $\beta = 1/4$ while
$S^{2}_+$ grows as $\sim t^{2/3}$ implying $\beta = 1/3$. Recalling
that $\chi =1/2$ for both modes, we have $z=2$ for the $-$ mode and $z
= 3/2$ for the $+$ mode.  There is a logarithmic factor
$(log(t))^{1/4}$, multiplying $t^{1/2}$ for the $-$wave as is apparent
from the flattening of the curve on dividing $S^{2}_-$ by
$t^{1/2}(log(t))^{1/4}$ (Fig. \ref{log3-}).
\figtwo{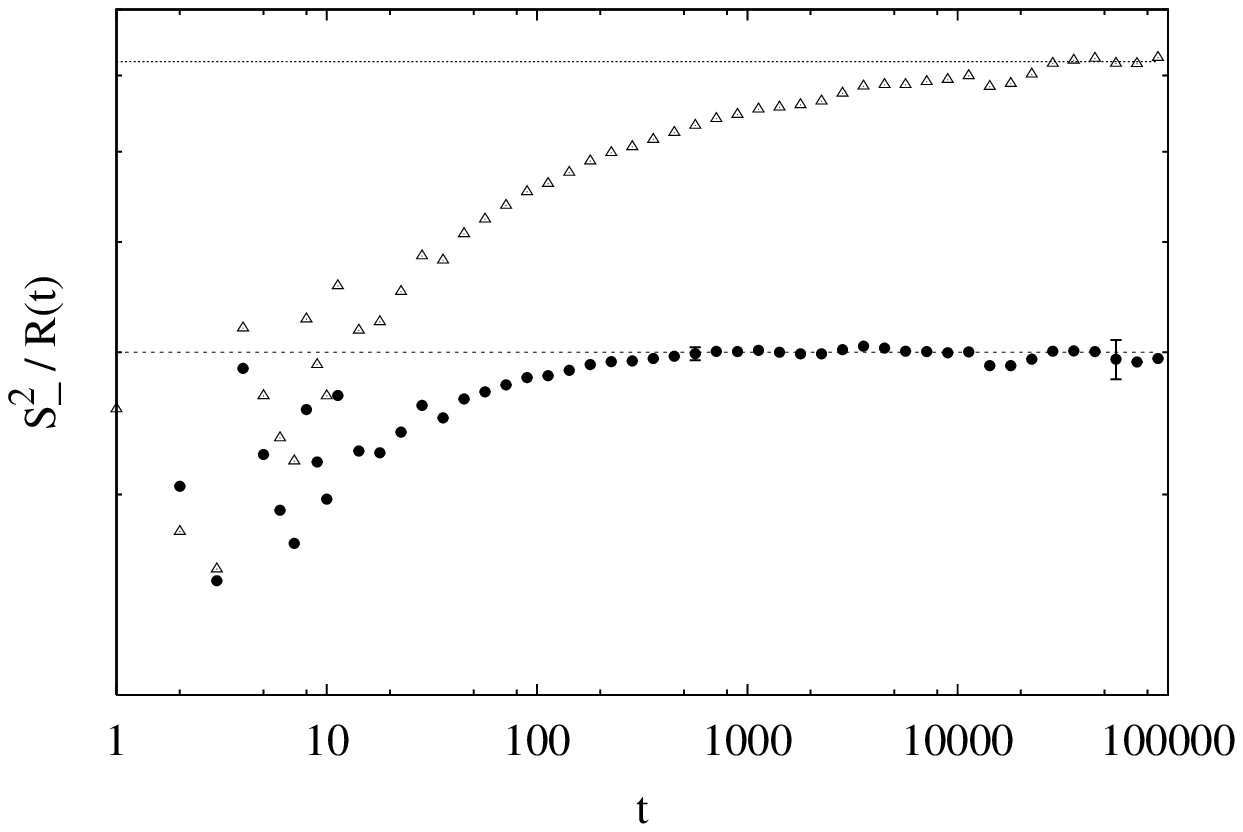}{Effect of rescaling  $S_-^2$ in Fig. \ref{corr0.33_0.33}
by $R(t) = t^{1/2}$ (upper curve) and $R(t)=t^{1/2}
(log(t))^{1/4}$ (lower curve). The flattening of the latter
confirms the presence of the logarithmic factor in $S^2_{-}$.}{log3-}
It is remarkable that although the waves are coupled 
to each other, two different dynamical exponents arise
in the same system, in conformity with our expectations on symmetry grounds.
This is the first instance we know of
 in which such a property arises in a fully coupled system in which neither
field evolves autonomously.
    
\section{Analytical demonstration of weak dynamic scaling} 
\label{analytwds} 
We argued in Sec. \ref{dynexplat} in the context of the model equations 
\ref{case3}, and showed by using Monte Carlo simulations, that there
are two {\em different} dynamical exponents 3/2 and 2 respectively,
for the two eigenmodes $h_+$ and $h_-$ in the model. In this section
we use self-consistent mode-coupling and renormalization group methods
to study the large-distance, long-time properties of the correlation
functions: We calculate the roughness exponents $\chi_{1,2}$ and the
dynamical exponents $z_{1,2}$ of the fields $h_{+,-}$. Our analytical
results agree with the previous numerical results.  For appropriately
chosen values for the densities of the particles in the lattice model
[see above eqn. (\ref{case3})], the continuum equations in the
center-of-mass frame take the form
%\begin{mathletters}
\begin{eqnarray}
\dot{h}_+-{B_o\over 2}\partial_xh_++{\lambda_1\over 2}(\partial_x h_+)^2+
{\lambda_2\over 2} (\partial_x h_-)^2&=&\nu_+\partial_{xx}h_++f_+,
\nonumber \\
%\label{sr1}
\dot{h}_-+{B_o\over 2}{\partial_x}h_-+\lambda_3(\partial_x h_+)(\partial_x h_-)
&=&\nu_-\partial_{xx}h_-+f_-,
%\label{sr2}
\label{sr}
\end{eqnarray}
%\label{sr}
%\end{mathletters}
where $B_o/2=c_+=-c_-,\,\lambda_1/2=3\kappa_2/\sqrt c,\,\lambda_2/2=-\kappa_2/2,
\,\lambda_3=2\kappa_2/\sqrt c$. We denote the dissipation coefficients by
$\nu_1$ and $\nu_2$ (we use two separate symbols in anticipation of
weak dynamic scaling). We also ignore the cubic nonlinearities for
simplicity.  The waves can be removed from either Eqs.(\ref{sr})
separately by comoving with the left and right going waves through the
Galilean shifts: $x\rightarrow x+{B_o\over 2}t,\,h_+\rightarrow h_+$ and
$x\rightarrow x-{B_o\over 2}t,\,h_-\rightarrow h_-$ respectively. In the
comoving frame of $h_+$ Eqs. (\ref{sr}) become
%\begin{mathletters}
\begin{eqnarray}
\dot{h}_++{\lambda_1\over 2}(\partial_x h_+)^2+
{\lambda_2\over 2} (\partial_x h_-)^2&=&\nu_+\partial_{xx}h_++f_+,\nonumber \\
%\label{srl1}
\dot{h}_-+B_o{\partial_x}h_-+\lambda_3(\partial_x h_+)(\partial_x h_-)
&=&\nu_-\partial_{xx}h_-+f_-,
%\label{srl2}
\label{srl}
\end{eqnarray}
%\label{srl}
%\end{mathletters}
whereas in the frame of $h_-$ Eqs.(\ref{sr}) reduce to
%\begin{mathletters}
\begin{eqnarray}
\dot{h}_+-B_o{\partial_x}h_h+{\lambda_1\over 2}(\partial_x h_+)^2+
{\lambda_2\over 2} (\partial_x h_-)^2&=&\nu_+\partial_{xx}h_++f_+,
\nonumber \\
%\label{srr1}
\dot{h}_-+\lambda_3(\partial_x h_+)(\partial_x h_-)
&=&\nu_-\partial_{xx}h_-+f_-,
%\label{srr2}
\label{srr}
\end{eqnarray}
%\label{srr}
%\end{mathletters}
The bare response functions of the two fields $h_+$ and $h_-$ are given by
\begin{equation}
G_o^+(k,\omega)={1\over i\omega +\nu_+k^2};\;\;\;\;G_o^-(k,\omega)=
{1\over i\omega-iB_ok+\nu_-k^2}
\end{equation}
in the right moving frame and 
\begin{equation}
G_o^+(k,\omega)={1\over i\omega +iB_ok+\nu_+k^2};\;\;\;\;G_o^-(k,\omega)=
{1\over i\omega+\nu_-k^2}
\end{equation}
in the left moving frame. The noise correlations in both the frames are given by
\begin{equation}
\langle f_{+,-}(0,0)f_{+,-}(x,t)\rangle=2D_{+,-}\delta(x)\delta(t).
\end{equation}
Note that there is no frame in which the drift terms in {\em both} equations 
vanish. Noise correlations however do not change with change in reference frames
as they are delta-correlated in time.

In this section we analytically calculate the dynamic and roughness 
exponents $\chi_i, \, z_i, 
\, i = +,-$ for the fields $h_i,\,i=+,-$ respectively, defined 
by $\langle [h_i(x,t) - h_i(0,0)]^2 \rangle = x^{\chi_i}g(x^{z_i}/t)$.
There are two limits in which their behaviour is well-understood
analytically.  First, in the {\em absence} of the {\em kinematic wave}
($B_o$) terms, Eqs. (\ref{sr}) reduce to the EK \cite{ertas1}
equations. Although the complete phase diagram of the latter is not
known, they do have a locally stable renormalization-group fixed point
belonging to the universality class of the KPZ equation \cite{KPZ},
with $\chi_+=\chi_-=1/2$ and $z_+ = z_- =3/2$. Secondly, in the {\em
absence} of the nonlinear terms coupling $h_+$ and $h_-$, the
kinematic wave terms can be removed separately in each equation by
opposite Galilean transformations, yielding scaling properties
independent of the wavespeed. In the previous section our Monte Carlo
results show that even when there is a nonlinear coupling, for 
particular densities, there is weak dynamic scaling: $z_+=3/2,~~
z_-=2$. Below, we offer an understanding of weak dynamic scaling in an
analytical framework.

The possible occurence of weak dynamical scaling in this problem, 
makes a renormalization-group treatment difficult: one cannot rescale 
time in two different ways for the two fields.  
We therefore adopt a self-consistent integral-equation approach, for which 
such weak scaling presents no difficulties.  We then discuss how to circumvent
the problems in applying an RG treatment.

\subsection{Self-consistent mode coupling calculation}
\label{modecoupling}
\subsubsection{KPZ equation in a moving frame}
\label{movkpz}
Before embarking on a calculation for our model, it is instructive to look at
the simpler case of
 a growing KPZ surface in a moving frame. The KPZ
equation will be supplemented by a linear first order gradient term:
\begin{equation}
{\partial h\over\partial t}+c{\partial h\over\partial x}+
\lambda({\partial h\over\partial x})^2=\nu\nabla^2 h+\eta,
\label{mkpz}
\end{equation}
with $\langle \eta(x,t)\eta(0,0)\rangle=2D\delta(x)\delta(t)$. We ask:
Does the wave affect the exponents? We know $\chi=1/2$ for the KPZ
equation in 1 dimension. Since $\chi$ gives the static probability
distribution, it is independent of reference frames. Due to the
exponent identity $\chi+z=2$ \cite{stanley}, it follows that $z=
3/2$. Thus in this particular case exponents are not affected by
waves. The question is, can we see this in a self-consistent or
renormalization-group calculation?  Recall that the dynamic exponent
is given by the width of the peak of the dynamic structure factor.
The location of the peak, as a result of the {\em wave} induced by our
transformation to a frame moving with speed $c$, is at $\omega=ck$.
The response and the correlation functions in that frame are thus
given by $G_o(k,\omega)={1\over i\omega-ick+\nu k^2}$ and
$C_o(k,\omega)={2D\over (\omega-ck)^2+\nu^2k^4}$ respectively. The
one-loop integral that produces a singular correction to $\nu$ is
\begin{equation}
I\sim -\int dqd\Omega {2D q(k-q)kq\over [(\Omega-ck)^2+\nu^2q^4]
[i(\omega-\Omega)-ic(k-q) +\nu (k-q)^2]}\Big|_{\omega=ck}
\sim -\int dq{2Dq(k-q)kq\over \nu q^2[\nu q^2+\nu (k-q)^2]},
\end{equation}
which is same as that one obtains in a standard calculation for the KPZ equation
in the rest frame. A similar expression holds for the correction to 
the correlation function. 
In both these integrals, the wave can be shifted away trivially.  
Thus we obtain $z=3/2$ and $\chi=1/2$, in 
agreement with our expectations. With this background we now present the
calculations for the exponents for our model. Interestingly, we 
will find that the effects of the waves cannot be trivially 
shifted away as they could in the simple example discussed above.  

\subsection{Self-consistent calculation for the coupled model}
\label{self}
The {\em mode coupling} approach to solving equations such as (\ref{sr})
 consists in obtaining diagrammatic perturbation expansions
for the renormalized propagator $G$ and correlation functions $C$ and resumming
these in such a way that all the {\em internal lines} are renormalized 
correlation  functions or propagators. This provides an exact solution for 
asymptotically small wave number $q$ when the vertex corrections vanish
for $q\to 0$ or due to some fundamental symmetry in the problem. In that
case the problem is reduced to solving nonlinear integral equations for
$G$ and $C$ whose order is the same as that of the nonlinearity in the
modified Langevin equation.
 
\begin{figure}[htb]
\epsfxsize=13cm
\centerline{\epsffile{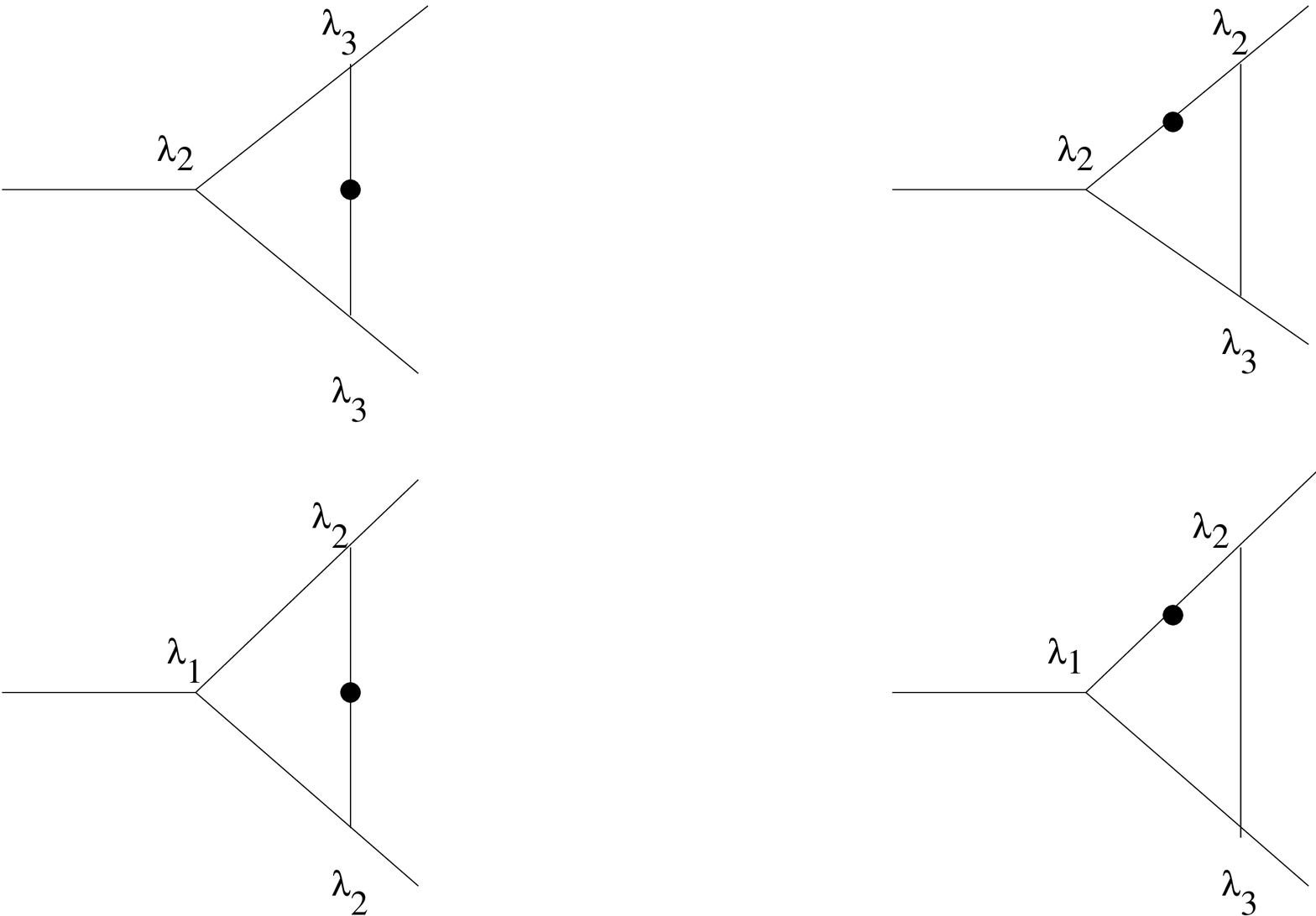}}
\caption{One-loop diagrammatic corrections to $\lambda_2$; A line indicates a
response function and a line with a small filled circle refers to a correlation
function.}
\label{l2}
\end{figure}

\begin{figure}[htb]
\epsfxsize=13cm
\centerline{\epsffile{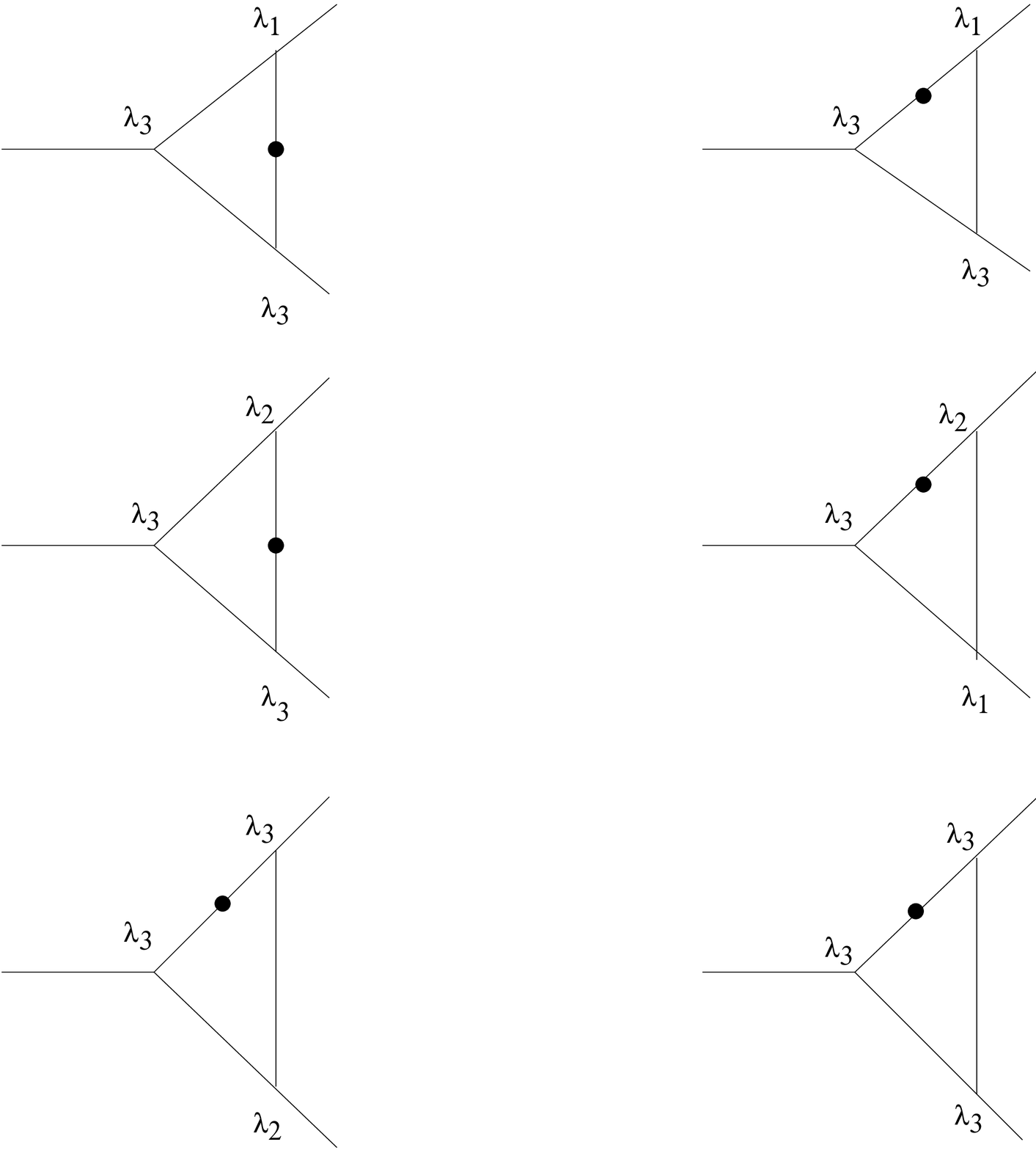}}
\caption{One-loop diagrammatic corrections to $\lambda_3$; A line indicates a
response function and a line with a small filled circle refers to a correlation
function.}
\label{l3}
\end{figure}

In a one-loop self-consistent modecoupling theory, one writes down
one-loop intergral equations for the response and correlation functions.
The basic assumption is that there are one-loop corrections which
diverge in the infrared limit.  
In a theory where there is no vertex renormalisation due to some Ward-Takahashi
identity (arising from some continuous symmetry of the equation of motion)
these equations are {\em exact}, because any higher loop corrections
can be incorporated within one-loop dressed response or correlation functions.
However, when there is no such symmetry of the system to prevent vertex
renormalisation the above assertion is not true. In such a situation,
one also has to write down a one-loop self-consistent equation for the
vertex, which now is to be solved simultaneously with the one-loop
equations for response and correlation functions. An example of this has
been worked out in \cite{abhik1}.
Here we work with Eqs.(\ref{sr}). There are no diagramatic corrections
to the $\lambda_1$ vertex at the one loop level. 
However, there are no such conditions on $\lambda_2$ and
$\lambda_3$. In our particular problem, however we show 
using bare response and correlation functions, that one loop
vertex corrections for $\lambda_2$ and $\lambda_3$ are all infrared finite.
Hence we ignore them in our calculations. In Figs. \ref{l2} and \ref{l3}
we show the one loop diagramatic corrections to $\lambda_2$ and $\lambda_3$
respectively. We first work in the comoving frame of $h_+$: In that frame
$C_+(k,\omega)$ has a peak at $\omega=B_ok$ and $C_-(k,\omega)$ is peaked
at $\omega=0$.
 Let us look at one of the diagrams very carefully (first diagram
in Fig.\ref{l2}):
\begin{equation}
I=ik\lambda_2\lambda_3^2/2 \int{d\Omega dqD_+q^2 q^2\over [(\Omega-B_oq)^2
+\nu_-^2q^4][\Omega^2+\nu_+q^4]}={2ik\lambda_2\lambda_3^2\over B_o^2}
({1\over \nu_+}+{1\over \nu_-})\int dq D_+.
\end{equation}
This has no infrared divergence. Similarly, all other diagrams in 
Figs.(\ref{l2}) 
and (\ref{l3}) are also finite in the infrared limit. We ignore all these finite
corrections for $\lambda_2$ and $\lambda_3$, i.e., we ignore vertex corrections
in our self-consistent analysis. We again justify this {\em a posteriori}, 
by showing that the vertex corrections remain finite in the 
self-consistent theory.  

The time-displaced correlation 
functions of the fields $h_+$ and $h_-$ are 
\begin{equation}
C_+(r,t)\equiv\langle h_+(0,0)h_+(r,t)\rangle,\;\;\;
C_-(r,t)\equiv\langle h_-(0,0)h_-(r,t)\rangle.
\end{equation}
The scaling forms for the correlation and response functions as a function of 
wavenumber $k$ and frequency $\omega$ are  
\begin{equation}
C_+(k,\omega)= k^{-1-2\chi_+-z_+}f_+(k^{z_+}/\omega),\;\;\;
C_-(k,\omega)= k^{-1-2\chi_--z_-}f_-(k^{z_-}/\omega),
\end{equation}
and
\begin{equation}
G_+(k,\omega)= k^{-z_+}g_+(k^{z_+}/\omega),\;\;\;
G_-(k,\omega)= k^{-z_-}g_-(k^{z_-}/\omega).
\end{equation}
Here, $z_+$ are $z_-$ are the dynamic exponents and $\chi_+$ and 
$\chi_-$ are the roughness exponents of the fields $h_+$ and $h_-$. Notice
that we have allowed the existence of two {\em different} dynamic
exponents. 
Since the fields are decoupled in the linearized theory,
there is no cross propagator.
The following one-loop diagrams contribute to the respective 
self-energies $\Sigma_+(k,\omega)$ and $\Sigma_-(k,\omega)$ (defined 
by $G_i^{-1} = G_{i0}^{-1} - \Sigma_i$, where $G_{i0}$ is the 
bare propagator for $h_i, \, i = +,-$) of the fields
$h_+$ and $h_-$.
\begin{figure}[htb]
\epsfxsize=14cm
\centerline{\epsffile{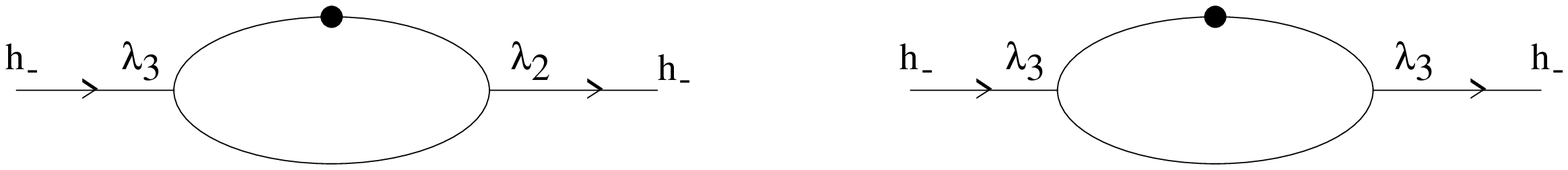}}
\label{hh2}
\caption{One-loop diagrams contributing to $\Sigma_-(k,\omega)$. A 
line refers to a response function and a line with a small filled circle
refers to a correlation function.}
\end{figure}
Notice that $G_{+0}^{-1}(k,\omega=0)=iB_ok+\nu_+k^2,\,G_{-0}^{-1}
=\nu_-k^2$.
A self-consistent calculation is required if one encounters infrared 
divergences in the bare perturbation theory. The following one loop
diagrams in Figs.10 and \ref{c2} contribute to $\Sigma_-(k,\omega)$
and $C_-(k,\omega)$ respectively. It is easy to see that none of these
diagrams diverge in a bare perturbation theory.
The  first diagram in Fig.10 has the form
\begin{eqnarray}
\Sigma_-(k,\omega)\sim \lambda_2\lambda_3k\int dq{D_-q^2(k-q)\over\nu_+ q^2
[\nu_+q^2+\nu_-(k-q)^2 -iB_o(k-q)]}&\sim &
\lambda_2\lambda_3 k\int dq {D_-q^2(k-q)[\nu_+q^2+\nu_-(k-q)^2]\over
\nu_-q^2[\{\nu_+q^2+\nu_-(k-q)^2\}^2+B_0^2(k-q)^2]}\nonumber \\ &\sim&
\lambda_2\lambda_3k\int dq{D_-(k-q)\over \nu_- B_0^2(k-q)^2}\sim {\rm finite},
\label{int1}
\end{eqnarray}
since in the long wavelength limit $B_0^2 (k-q)^2$ dominates over 
$\{\nu_+ q^2+\nu_-(k-q)^2\}^2$.
We have considered only the real part of $\Sigma_-(k)$ (only this part
will renormalise $\nu_-$). However this does not produce any infra-red
singular correction to $\nu_2$ and hence we ignore it. Note that it is the 
presence of the kinematic waves  which makes this diagrammatic contribution
infrared finite and thus ignorable.
The second diagram in Fig. (10) also has a similar finite form and is ignored
again.  
\begin{figure}[htb]
\epsfxsize=10cm
\centerline{\epsffile{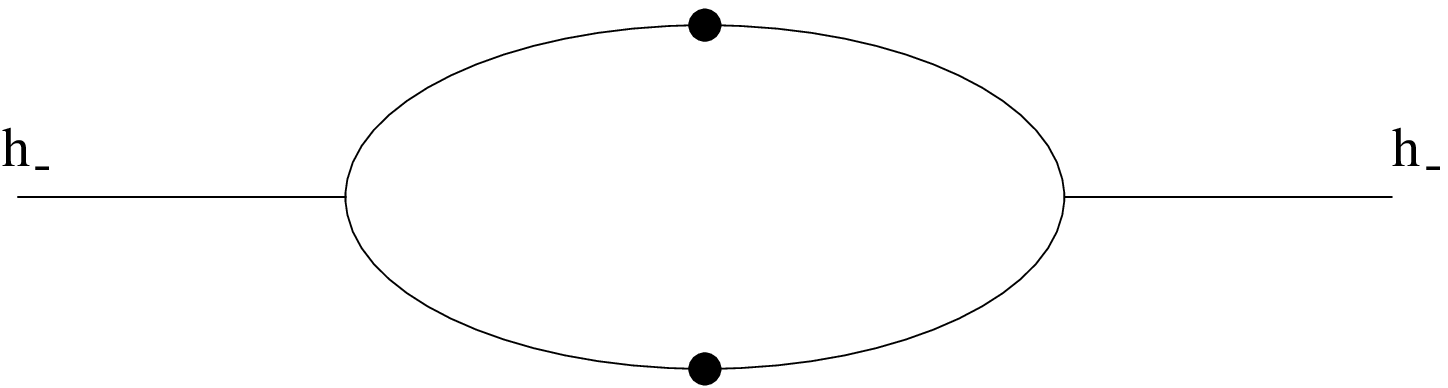}}
\caption{One-loop contribution to $C_2(k,\omega)$. A line refers to a response
function and a line with a small filled circle refers to a correlation function
.}
\label{c2}
\end{figure}
In Fig.\ref{c2} the  diagram has the form
\begin{equation}
C_-(k,\omega)\sim {1\over k^{4}}\int {dq~~q^2(k-q)^2\over
q^2(k-q)^2[\nu_-q^2+\nu_+(k-q)^2-iB_o(k-q)]}.
\label{int2}
\end{equation}
It is easy to see that this is not infrared divergent. So we
ignore all corrections to $C_{-}(k,\omega)$. We
immediately obtain $z_-=2,\,\chi_-=1/2$. Note that the presence 
of the wave term $B_0k$ in the inverse propagator was again crucial 
to this analysis. 
We see from the foregoing analysis that 
\begin{equation}
G_{-}^{-1}(k,\omega)\sim {i\omega+k^2},\,C_-(k,\omega=0)\sim 
{1\over \omega^2+k^4}.
\end{equation}

\begin{figure}[htb]
\epsfxsize=14cm
\centerline{\epsffile{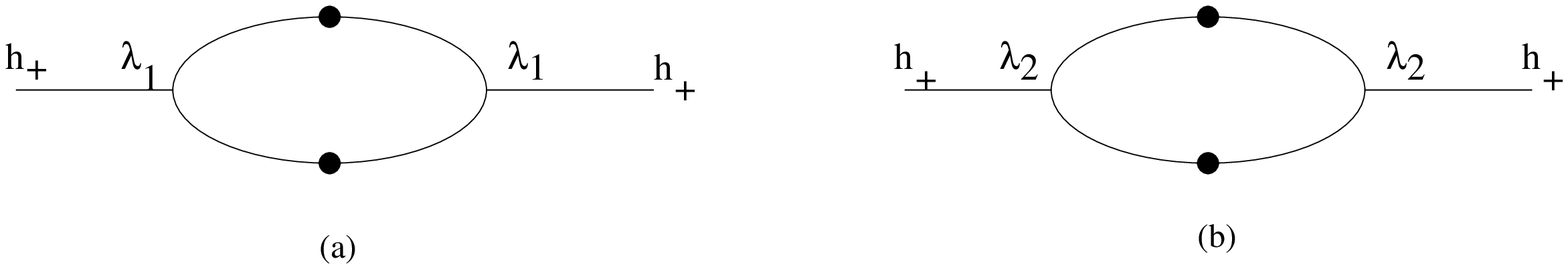}}
\caption{One-loop contribution to $C_+(k,\omega)$. A line refers 
to a response function and a line with a small filled circle refers 
to a correlation function.}
\label{hh1}
\end{figure}

\begin{figure}[htb]
\epsfxsize=15cm
\centerline{\epsffile{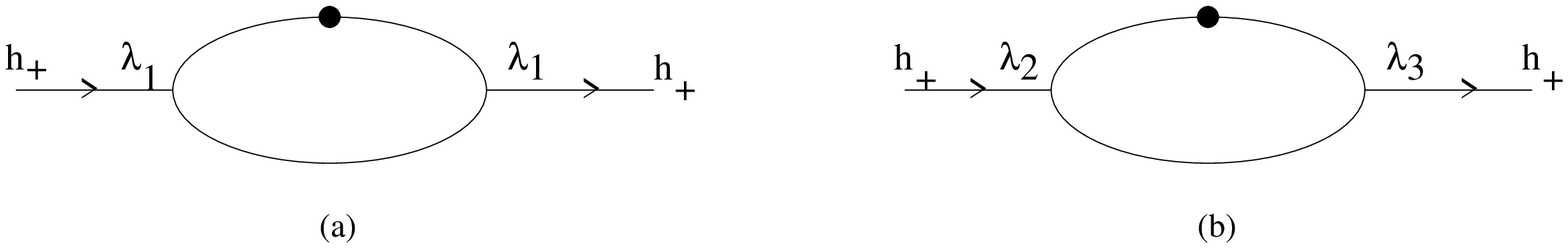}}
\caption{One-loop contribution to $\Sigma_+(k,\omega)$. A line refers
to a response function and a line with a small filled circle refers
to a correlation function.}
\label{C1}
\end{figure}

We now calculate the exponents of $h_+$. As stated earlier 
we ignore corrections to
$\lambda_2$ and $\lambda_3$ as they are all finite. There are however,
diverging one-loop corrections to both $C_+$ and $\Sigma_+$ which are shown in
Fig.\ref{hh1} and Fig.\ref{C1}.
In both Figs. \ref{hh1} and \ref{C1} diagrams (a) comes from the KPZ 
nonlinearity. If these diagrams were the only diagrams, one would have obtained
KPZ exponents:$z_+=3/2,\,\chi_+=1/2$. We however notice that the diagrams (b)
in Figs.\ref{hh1} and \ref{C1} are {\em as} strongly infrared (IR) divergent as
diagrams (a). Let us examine diagram (b) in Fig.\ref{hh1} in detail.  The
integral is given by (using the fact that $\chi_-=1/2,\,z_-=2$)
\begin{eqnarray}
\label{diagb}
I&\sim& k\int d\Omega dq{q^2(k-q)\over (\Omega^2+\nu_-^2q^4)[-i\Omega+iB_ok
+\nu_-(k-q)^2]}\sim k\int dq{(k-q)[\nu_-q^2+\nu_-(k-q)^2]\over\nu_- 
[B_o^2 k^2+\{\nu_- q^2+\nu_-(k-q)^2 \}^2]}\nonumber \\
&\sim& k\int dq{(k/2-q)\nu_- q^2\over\nu_-[B_o^2k^2+4\nu_-^2q^4]}
\sim {k^2\over \nu_-}\int_{\sqrt k}{dq\over q^2}[1+O(k^2/q^4)+..]\sim k^{3/2}.
\end{eqnarray}
Note that the presence of the kinematic wave was crucial again in this 
evaluation: The integral in (\ref{diagb}) diverges as $\int dq / q^2$ 
if the external wavenumber 
$k$ is set to zero. At nonzero $k$, the integral is controlled 
by the presence of the wave term $B_0^2 k^2$ in the denominator, 
and scales as $k^{-1/2}$, not as $k^{-1}$ as might have naively been expected.  
Thus $z_+=3/2$ is unaltered by the second diagram. Similarly we consider
diagram (b) of Fig.\ref{C1} (using $z_+=3/2$):
\begin{equation}
I\sim {1\over k^3}\int d\Omega dq{q^4\over (\Omega^2+\nu_-^2 q^4)
[(\Omega-B_ok)^2+\nu_-^2q^4]}\sim {1\over i\nu_-k^3B_ok}\int dq q^2[{1\over
B_ok-i\nu_-q^2}-{1\over B_ok-i\nu_-q^2}]\sim k^{-7/2},
\end{equation}
which is as divergent as the diagram (a) in Fig.\ref{C1}. Thus
$\chi_+=1/2$ also remains unaltered. It is easy to see that with these 
self-consistent
response and correlation functions, one-loop corrections to $\lambda_2,
\lambda_3$ do not diverge. Hence ignoring $\lambda_2,\lambda_3$ from our
selfconsistent calculation is justified.

So far we have worked in the comoving frame of $h_-$, i.e, with Eqs.
(\ref{srr}). Let us now go to the comoving frame of
$h_+$: We work with Eqs. (\ref{srl}). The relevant  diagrams
are same as given in Figs. \ref{hh1} and \ref{C1}. 
It is easy to carry out a self-consistent analysis again on 
Eqs. (\ref{srl}). The only difference  is
that now $C_+(k,\omega)$ is peaked at $\omega=0$
whereas $C_-(k,\omega)$ is peaked at $\omega=-B_ok$.  
Here again there is no singular correction to the reponse and the correlation
functions of the field $h_-$. Thus $z_-=2$ and $\chi_-=1/2$. There are
however diverging corrections to $\nu_1$ and $D_1$ which are identical
to those obtained in our calculations in the comoving frame of $h_-$.
Reassuringly, we find again {\em same} exponents, i.e. $z_+=3/2,\,\chi_+=1/2$, 
as we should.
Thus the effective equations in the left and right going frames are
%\begin{mathletters}
\begin{eqnarray}
{\partial h_+(k)\over\partial t}+iB_okh_+&=&\nu_+k^{3/2}h_+(k,t)+....,
\nonumber \\
%\label{rer1}
{\partial h_-(k)\over\partial t}&=&\nu_-k^{2}h_-(k,t)+....
\label{rer}
\end{eqnarray}
%\label{rer}
%\end{mathletters}
and
%\begin{mathletters}
\begin{eqnarray}
{\partial h_+(k)\over\partial t}&=&\nu_+k^{3/2}h_+(k,t)+....,\nonumber\\
%\label{rel1}
{\partial h_-(k)\over\partial t}-iB_okh_-&=&\nu_-k^{2}h_-(k,t)+....
\label{rel}
\end{eqnarray}
%\label{rel}
%\end{mathletters}
The renormalized equations (\ref{rer}) and (\ref{rel})
are thus connected by a Galilean transformation like the bare equations
(\ref{srr}), and (\ref{srl}), and the exponents are 
frame-independent.

\subsection{Renormalization group analysis}
\label{rg}
In Section \ref{self} we have shown how a self-consistent mode coupling 
treatment
of our model for certain parameter values leads to weak dynamical 
scaling, i.e., {\em distinct}
dynamical exponents for the two fields. In this Section we recast these  
results within a perturbative dynamical renormalization-group (DRG) framework, and show 
how the difficulties posed for the DRG by weak dynamical scaling can be 
overcome.  

The diagrams are same as shown in 
Section \ref{self}. We work with Eqs. (\ref{srl}). As usual,
the bare diagrams for $\nu_-,\,D_-,\,\lambda_2$ and $\lambda_3$ do not diverge.
The diagram (a) in Fig.\ref{hh1} diverges as $\sim {D_+\lambda_1^2\over\nu_1^3}
k^2\int {dq\over q^2}\sim {D_+\lambda_1^2\over\nu_1^3}k^2\times {1\over k},$
whereas the diagram (b) in Fig.\ref{hh1} diverges as
$\lambda_2\lambda_3D_-k^2\int {dq\over iB_ok+2\nu_-q^2}\sim
\lambda_2\lambda_3D_-k^2\times {1\over\sqrt k}$. Thus diagram (b) can be
ignored in comparison with (a) in the small $k$ limit. Similarly, diagram
(b) in Fig.\ref{C1} is less divergent than the diagram (a). Thus all relevant
diagrams generating dominant singular corrections to the bare response
and correlation functions of the field $h_+$ are of KPZ type (with 
identical symmetry factors). 

Starting with a cutoff wavenumber $\Lambda$, eliminating modes 
with wavenumbers between $\Lambda e^{-\delta l}$ and $\Lambda$, 
rescaling so that the cutoff wavenumber is once more $\Lambda$, 
and passing to the limit $\delta l \to 0$, we obtain 
the differential recursion relations 
%\begin{mathletters}
\begin{eqnarray}
{d\nu_+\over dl}&=&\nu_+[z-2+{g\over 4}],\nonumber \\
{dD_+\over dl}&=&D_+[z-1-2\chi_++{g\over 4}],
\label{fl12}
\end{eqnarray}
and
\begin{eqnarray}
{d\nu_-\over dl}&=&\nu_-[z-2],\nonumber \\
{dD_-\over dl}&=&D_-[z-1-2\chi_-],
\label{fl34}
\end{eqnarray}
where $\nu_+,\nu_-,D_+$ and $D_-$ are now functions of $l$, 
and $g \equiv D_+ \lambda_1^2 / \nu_+^3$ is the dimensionless 
coupling constant.  
By using $g=2$ at the stable RG fixed point \cite{stanley}, we obtain
\begin{mathletters}
\begin{eqnarray}
{d\nu_+\over dl}&=&\nu_+[z-3/2],\\
{d\nu_-\over dl}&=&\nu_-[z-2],\\
{dD_+\over dl}&=&D_+[z-1/2-2\chi_+],\\
{dD_-\over dl}&=&D_-[z-1-2\chi_-].
\end{eqnarray}
\end{mathletters}
Now given the exponents $z,\chi_+,\chi_-$, the equations of motion (\ref{srl})
are 
supposed to be invariant under the scale transformations $x\to bx,\,t\to b^zt,
\,h_+\to b^{\chi_+}h_+,\,h_-\to b^{\chi_-}h_-$. From Eqs. (\ref{fl12}) 
we get $z=3/2,\chi_+=1/2$ whereas Eqs. (\ref{fl34}) 
give $z=2,\chi_-=1/2$. Which value of $z$ should we choose? The choice of
$z=3/2$ keeps $\nu_1$ fixed under rescaling, but $\nu_2(l)\sim e^{-l/2}$ as the
RG fixed point is approached (remember $e^l$ is like a length scale). 
On the other hand
if we choose $z=2$ then $\nu_2$ is fixed under rescaling but $\nu_1\sim 
e^{l/2}$ as the fixed point is approached. Moreover, $z=2$ implies 
$\chi_-=1/2$ giving
$C_-(x,0)\sim x$ in agreement with the previously obtained results, and
$\chi_+=1/4$ suggesting $C_+(x,0)\sim x^{1/2}$, which, of course, is wrong.
Similarly, the choice $z=3/2$ gives correct spatial dependence for $C_+(x,0)
\sim x$, but gives an incorrect result for $C_-(x,0)$.

To resolve this difficulty, let us first study the seemingly trivial 
case of two totally decoupled fields:    
\begin{mathletters}
\begin{eqnarray}
{\partial \phi_1\over\partial t}&=&\mu_1{\partial^2 \phi_1\over\partial x^2}
+\eta_1,\\
{\partial \phi_2\over\partial t}&=&-\mu_2{\partial^2 \phi_2\over\partial x^4}
+\eta_2,
\end{eqnarray}
\end{mathletters}
with $\langle \eta_i(0,0)\eta_j(x,t)\rangle=2\gamma_i \delta_{ij} 
\delta(x)\delta(0)$ for $i = 1, 2$ and no sum on repeated indices. 
These are linear equations and hence exponents can be found exactly. In
particular we know
\begin{mathletters}
\begin{eqnarray}
\langle\phi_1(x,t)\phi_1(0,0)\rangle&\sim& xf_1(x^2/t),\\
\langle\phi_2(x,t)\phi_1(0,0)\rangle&\sim& x^{3}f_2(x^4/t),
\end{eqnarray}
\end{mathletters}
i.e., we have $\chi_1=1/2,\,\chi_2=3/2,\,z_1=2,\,z_2=4$.
There are obviously no diagramatic corrections to any of the parameters. 
Here the flow equations are
\begin{mathletters}
\begin{eqnarray}
{d\mu_1\over dl}&=&\mu_1[z-2];\\
\label{ll1}
{d\gamma_1\over dl}&=&\gamma_1[z-1-2\chi_1];\\
\label{ll2}
{d\mu_2\over dl}&=&\mu_2[z-4];\\
\label{ll3}
{d\gamma_2\over dl}&=&\gamma_2[z-1-2\chi_2].
\label{ll4}
\end{eqnarray}
\end{mathletters}
A choice of $z=2$ and $\chi_1=1/2$ keeps $\mu_1$ and $\gamma_1$ fixed. This
immediately tells us that equal-time correlation function of the field
$\phi_1$ scales as $Ax^{2\chi_1}$. The coefficient $A$, a function of $\gamma_1$
and $\mu_1$ is scale independent. However a na\"{i}ve use of this value
of $z$ (i.e.,$z=2$) leads to wrong conclusion about the spatial dependence
of  the equal time correlation function of the field of $\phi_2$: $z=2$
gives $\chi_1=1/2$ suggesting $\langle \phi_2(0,0)\phi_2(x,0)\rangle
\sim x$ which is wrong (the correct dependence is $x^3$ which
is known {\em exactly}). The reason is that the choice $z=2$
makes the solutions of the flow equations for $\mu_2$ scale-dependent: 
$\mu_2\sim e^{-2l}\sim k^2$. However,
one can still extract the correct behaviour of the correlation function 
of $\phi_2$ in the above example if one accounts for the 
scale dependence $\mu_2$ explicitly: An explicit construction gives
\begin{equation}
\langle \phi_2(k,\omega)\phi_2(-k,\omega)\rangle\equiv {D_2\over\mu_2
k^{1+2\chi_1}}[{1\over i\omega +\mu_2k^z}+{1\over -i\omega +\mu_2k^z}],
\end{equation}
which gives (by using scale dependent $\mu_2$)
$\langle \phi_1(0,0)\phi_2(x,0)\rangle\sim x^3$ which is the 
correct answer! This suggests that to make sense of out the RG flow 
equations in presence of different dynamic scalings in a coupled system,
one has to take care of the scale dependent diffusion coefficients 
while calculating the spatio-temporal behaviour of the 
correlation functions correctly. Let us review our RG results 
obtained from the flow equations (\ref{fl12}) and (\ref{fl34}) in 
view of our previous analysis: We  choose $z=2,\,\chi_-=1/2$ which gives
$\chi_+=1/4$ and makes $\nu_1\sim k^{-1/2}$. Using this
scale dependent coefficient we correctly obtain
\begin{equation}
C_+(x,t)=xf_-[x^{3/2}/t].
\end{equation}
This suggests that an RG treatment, suitably modified, can be applied 
successfully to a problem of weak dynamic
scaling provided scale dependent dissipation coefficients are taken into
account while constructing the correlation  functions.

\subsection{Strong dynamical scaling}
\label{strong}
In Section \ref{expfrommcsim} we have seen that for model equations 
 \ref{case2} both the dynamic exponents turn out to be 3/2 (i.e., KPZ-like).
This is easy to understand analytically: In this case both the equations
have KPZ-like nonlinearities in addition to the non-KPZ ones. However,
it is easy to see that for both the fields, due to the presence of the waves,
in any frame, non-KPZ diagrams are less singular than the 
corresponsing KPZ-like diagrams. Thus the dominant singular corrections to
the response and correlation functions for both the fields are KPZ-like,
making $\chi_+=\chi_-=1/2$ and $z_+=z_-=3/2$.

\section{summary}
\label{summary} 
This paper is a study of the nature of spatiotemporal correlations 
in a coupled-field driven 
diffusive model (the LR model \cite{rlsr,lahiri2}),  
in the phase in which it displays kinematic waves.  
Our results include a demonstration of pairwise balance for certain parameter 
values and, hence, a proof that the steady state has a product measure 
in that range of parameters. Most important,  
we have been able to show that the fields corresponding to 
the two eigenmodes of the linearized version of the model were characterized by 
two different dynamic exponents, although the fields themselves are 
(nonlinearly) coupled. This is the first demonstration of such 
{\em weak dynamical scaling} in a model with two fully 
coupled fields (as distinct from the model of \cite{ertas1,bara1,abhik1}, 
where such an effective decoupling was found 
only in semi-autonomously coupled systems in which one of the fields
evolve independently).  
We have been able to show this numerically, through Monte Carlo simulations
on a lattice
model, and analytically, using self-consistent perturbative calculations as 
well as symmetry arguments, in the corresponding continuum stochastic 
PDEs. 
Outisde this special subspace of parameter space, the model exhibits 
normal, strong dynamic scaling. We also discuss and largely resolve 
the technical difficulties  
in applying the dynamical renormalization group when weak dynamic scaling 
prevails. 

\section{Acknowledgement}
We thank Goutam Tripathy for valuable discussions. 


\begin{references}
\bibitem[\P]{bydib}Present adress:Martin Fisher School of Physics,
        Brandeis University, Mailstop 057
        Waltham, Massachusetts 02454-9110, USA.\\
email:dibyendu@octane.cc.brandeis.edu
\bibitem[\ast]{bypur}email:abhik@physics.iisc.ernet.in
\bibitem[\S]{bybarma}email:barma@theory.tifr.res.in
\bibitem[\ast\ast]{byjnc}Also at Jawaharlal Nehru Centre for Advanced 
Scientific Research, Bangalore, India.\\
email:sriram@physics.iisc.ernet.in
\bibitem{lighthill}M.J. Lighthill and G.B. Whitham, Proc. R. Soc. A {\bf 229},
281, 317 (1955).
\bibitem{epjb}A. Basu, J.K.Bhattacharjee, and S.Ramaswamy, {\em Eur. Phys. J B}
{\bf 9}, 425 (1999).
\bibitem{ertas1}D. Erta\c{s} and M. Kardar, {\em Phys. Rev. E}{\bf 48},
1228 (1993).
\bibitem{rlsr}R. Lahiri and S. Ramaswamy, {\em Phys. Rev. Lett.} {\bf 79},
1150 (1997).        
\bibitem{lahiri2}R. Lahiri, M. Barma and S. Ramaswamy, {\em Phys. Rev. E}
{\bf 61}, 1648 (2000).
\bibitem{jnurev}S. Ramaswamy, M. Barma, D. Das and A. Basu, preprint 
TIFR/TH/00-58,
\bibitem{ddmb}D. Das and M. Barma, Phys. Rev. Lett.{\bf 85}, 1602 (2000)
\bibitem{pairwise}G. M. Sch\"{u}tz. R. Ramaswamy and M. Barma, J. Phys. A:
Math. Gen. {\bf 29} 837 (1996).
\bibitem{kampen}N.G. van Kampen, {\it Stochastic Processes in Physics and
Chemistry} (North Holland, Amsterdam, 1992).
\bibitem{stanley} A.-L. Barab\'{a}si and H.E. Stanley, {\it Fractal Concepts
in Surface Growth}, (Cambridge, New York, 1995).
\bibitem{EW}{S. F. Edwards and D.R. Wilkinson, Proc.
R. Soc. London, Ser. A {\bf 381}, 17 (1982).}
\bibitem{spohn1}P. Devillard and H. Spohn, J. Stat. Phys. {\bf 66}, 1089
(1992).
\bibitem{spohn2}M. Paczuski, M. Barma, S.N. Majumdar and T. Hwa, Phys.
Rev. Lett. {\bf 69}, 2735 (1992).
\bibitem{spohn3}P.M. Binder, M. Paczuski and M. Barma, Phys. Rev. E {\bf 49}
1174 (1994).
\bibitem{spohn4}{B. Subramanian, G.T. Barkema, J.L. Lebowitz and
E.R. Speer, J. Phys. A: Math. Gen. {\bf 29} 7475 (1996).}
\bibitem{KPZ}{M. Kardar, G. Parisi and Y.C. Zhang, Phys. Rev. Lett. {\bf 62},
89 (1986)}.
\bibitem{kardar}M. Kardar, cond-mat/9704172.
\bibitem{bara1}A.-L. Barab\'{a}si, Phys. Rev. A {\bf 46}, R2977 (1992).
\bibitem{goutam} G. Tripathy and M. Barma, Phys. Rev. E{\bf 58}, 1911 (1998).
\bibitem{abhik1}A.K. Chattopadhyay, A. Basu and J.K. Bhattacharjee, Phys.
Rev. E {\bf 61}, 2086 (2000).
\end{references}
\end{document}